\documentclass[12pt]{article}

\usepackage[top=80pt,bottom=85pt,left=60pt,right=60pt]{geometry}
\usepackage{amssymb,amsmath,xypic}
\usepackage{slashed}
\usepackage{graphicx,color,xcolor,subfigure}
\usepackage{float}
\usepackage{sectsty}
\usepackage{cite}
\usepackage[debug,pageanchor=false]{hyperref}
\definecolor{link}{rgb}{.8,.15,.1}
\hypersetup{colorlinks=true,linkcolor=link,citecolor=link,urlcolor=link,linktocpage}

 \allowdisplaybreaks

\makeatletter
\@addtoreset{equation}{section}
\makeatother

\newcommand{\beq}{\begin{equation}}
\newcommand{\eeq}{\end{equation}}
\newcommand{\bea}{\begin{eqnarray}}
\newcommand{\eea}{\end{eqnarray}}
\newcommand{\nn}{\nonumber}

\def\white1{\textcolor[rgb]{0.98,0.98,0.98}}
\usepackage{rotating}

\begin{document}

\begin{titlepage}

\begin{center}

\vskip .5in 
\noindent

{\Large \bf{ AdS$_3$ vacua realising $\mathfrak{osp}(n|2)$ superconformal symmetry  }}

		\bigskip\medskip
 Niall T. Macpherson$^{a}$\footnote{macphersonniall@uniovi.es}, Anayeli Ramirez$^{b}$\footnote{Anayeli.Ramirez@mib.infn.it} \\

\bigskip\medskip
{\small 
	
	$a$: Department of Physics, University of Oviedo,\\ 
	Avda. Federico Garcia Lorca s/n, 33007 Oviedo
\\
	and
\\
	 Instituto Universitario de Ciencias y Tecnolog\'ias Espaciales de Asturias (ICTEA),\\ 
	Calle de la Independencia 13, 33004 Oviedo, Spain }\vskip 3mm

	$b$: Dipartimento di Fisica, Universit\`a di Milano--Bicocca, \\ Piazza della Scienza 3, I-20126 Milano, Italy \\ and \\ INFN, sezione di Milano--Bicocca

		\vskip 1.5cm 
		\vskip .9cm 
		{\bf Abstract }

		\vskip .1in
	
	\noindent 
We consider ${\cal N}=(n,0)$ supersymmetric AdS$_3$ vacua of type II supergravity realising the superconformal algebra $\mathfrak{osp}(n|2)$ for $n>4$. For the cases $n=6$ and $n=5$, one can realise these algebras on backgrounds that decompose as foliations of AdS$_3\times \mathbb{CP}^3$ ( squashed $\mathbb{CP}^3$ for $n=5$) over an interval. We classify such solutions with bi-spinor techniques and find the local form of each of them: They only exist in (massive) IIA and are defined locally in terms of an order 3 polynomial $h$ similar to the AdS$_7$ vacua of (massive) IIA.  Many distinct local solutions exist for different tunings of $h$ that give rise to bounded  (or semi infinite) intervals bounded by physical behaviour. We show that it is possible to glue these local solutions together by placing D8 branes in the interior of the interval without breaking supersymmetry, which expands the possibilities for global solutions immensely. We illustrate this point with some simple examples. Finally we also show that AdS$_3$ vacua for $n=7,8$ only exist in $d=11$ supergravity and are all locally AdS$_4\times$S$^7$.

\end{center}
\vskip .1in

\noindent

\noindent

\vfill
\eject

\end{titlepage}

\tableofcontents

\section{Introduction and summary}

Warped AdS$_3$ solutions of  supergravity in 10 and 11 dimensions, ``AdS$_3$ string vacua'', play an important role in string theory in a wide variety of contexts. AdS$_3$ appears in the near horizon limit of black-strings solution, so the embedding of such solutions into higher dimensions enables one to employ string theory  to count the micro states making up the Bekenstein--Hawking entropy a la Strominger--Vafa \cite{Strominger:1996sh}. Through the AdS-CFT correspondence they are dual to the strong coupling limit of CFTs in 2 dimensions. This avatar of the correspondence promises to be the most fruitful as more powerful techniques are available to probe CFT$_2$s and there is better understanding of how to quantise strings on AdS$_3$ than in higher dimensional cases.  AdS$_3$ vacua also commonly appear in duals to compactifications of CFT$_4$ on Riemann surfaces \cite{Maldacena:2000mw,Ferrero:2020laf,Boido:2021szx,Suh:2021ifj,Couzens:2021tnv,Arav:2022lzo,
Amariti:2023mpg,Suh:2023xse},  a topic of rekindled interest in recent years with improved understanding of compactifications on surfaces of non-constant curvature such as spindles. Some other venues in which AdS$_3$ vacua have played a prominent role are geometric duals to  c-extremisation \cite{Couzens:2018wnk,Couzens:2022agr} and dual descriptions of surface defects in higher dimensional CFTs \cite{DHoker:2008rje,DHoker:2009lky,Faedo:2020nol,Lozano:2022ouq,Anabalon:2022fti}. 

Given the above listed wealth of applications, a broad effort towards classifying supersymmetric AdS$_3$ vacua is clearly well motivated, but at this time many gaps remain. Generically such AdS$_{d+1}$ vacua can support the same superconformal algebras as CFTs in $d$ dimensions. The possible $d=2$ superconformal algebras are far more numerous than their higher dimensional counterparts, which partially accounts for these gaps. For comparison $d>2$ the possible (simple) superconformal algebras typically\footnote{$d=5$ is an exception with only one possibility, $\mathfrak{f}(4)$} come in series depending on a parameter which varies as the number of super charges increase; for instance in $d=3$ one has $\mathfrak{osp}(n|4)$ for CFTs preserving ${\cal N}=(n,0)$ supersymmetry, where $n=1,...,8$. CFTs in $d=2$ buck this trend, being consistent with several such series as well as isolated examples such as $\mathfrak{f}(4)$ and $\mathfrak{g}(3)$ - see \cite{Fradkin:1992bz} for a classification of these algebras and \cite{Beck:2017wpm} for those that can be supported by string vacua. The focus of this work will be  AdS$_3$ vacua supporting the algebra $\mathfrak{osp}(n|2)$  (the $d=2$ analogue of the $d=3$ algebra).\\
~\\
The ${\cal N}=(n,0)$ superconformal algebra $\mathfrak{osp}(n|2)$ for arbitrary $n$ was first derived, independently, in 
\cite{Bershadsky:1986ms} and \cite{Knizhnik:1986wc} -- they are characterised by an $\mathfrak{so}(n)$ R-symmetry with supercurrents transforming in the fundamental representation and central charge
\beq\label{eq:cftcentralcharge}
c= \frac{k}{2}\frac{n^2+6k-10}{k+n-3}.
\eeq
 A free field relation was presented in \cite{Mathieu:1988aa} (see also \cite{Khviengia:1995wp}) in terms of a free scalar, $n$ real fermions, and an SO($n$) current algebra of level $k-1$.
There are in fact many examples of AdS$_3$ vacua realising $\mathfrak{osp}(n|2)$ for $n=1,2$ as these are the unique ways to  realise  $(1,0) $ and  $(2,0) $ superconformal symmetries -- see for instance respectively\cite{Martelli:2003ki,Tsimpis:2005kj,Babalic:2014fua,Babalic:2014dea,Dibitetto:2018ftj,Passias:2019rga,Passias:2020ubv,Farakos:2020phe,Macpherson:2021lbr,VanHemelryck:2022ynr} and 
\cite{Kim:2005ez,Gauntlett:2006af,Gauntlett:2006qw,Gauntlett:2006ns,Donos:2008ug,Donos:2008hd,Couzens:2017nnr,Eberhardt:2017uup,Couzens:2019iog,Couzens:2019mkh,Couzens:2022agr,Ashmore:2022kho}
. Similarly $n=3$ is unique for  ${\cal N}=(3,0)$, examples  are more sparse 
\cite{Figueras:2007cn,Legramandi:2019xqd,Eberhardt:2018sce}, but this is likely not a reflection of their actual rarity. The case of $n=4$ is in fact a degenerate case of the large ${\cal N}=(4,0)$ superconformal algebra $\mathfrak{d}(2,1,\alpha)$, where the continuous parameter is tuned to $\alpha=1$ -- examples of vacua allowing such a tuning include \cite{deBoer:1999gea,Bachas:2013vza,Kelekci:2016uqv,Macpherson:2018mif}, there is also a Janus solution preserving $\mathfrak{osp}(4|2)\oplus \mathfrak{osp}(n|2) $ specifically in \cite{DHoker:2009lky}. The case of $n=8$ was addressed in \cite{Legramandi:2020txf} where it was shown that the only solution is the embedding of AdS$_3$ into AdS$_4\times$S$^7$. The status of AdS$_3$ vacua realising $\mathfrak{osp}(n|2)$ for $n=5,6,7$ has been up to this time unknown -- a main aim of this work is to fill in this gap, we will now explain the broad strokes of how we approach this problem.

For the case of $\mathfrak{osp}(7|2)$, with a little group theory \cite{Legramandi:2020txf}, it is not hard to establish that the required $\mathfrak{so}(7)$ R-symmetry can only be realised geometrically on the co-set SO(7)/G$_2$. The metric on this space is simply the round 7-sphere, which possesses an SO(8) isometry, but the co-set also comes equipped with a weak G$_2$ structure with associated 3 and 4-forms that are invariant under SO(7), but charged under SO(8)/SO(7). If these forms appear in the fluxes of a solution then only SO(7) is preserved. Our results prove that all such solutions are locally AdS$_4\times\text{S}^7$.

To realise the requisite $\mathfrak{so}(6)$ R-symmetry of $\mathfrak{osp}(6|2)$ one might naively consider including a 5-sphere in solutions, however this supports Killing spinors in the \textbf{4} of SU(4), which is not the representation associated to the desired algebra. A space supporting both the correct isometry and spinors transforming in its fundamental representation is of course round $\mathbb{CP}^3$, as famously exemplified by the AdS$_4\times\mathbb{CP}^3 $ vacua of type IIA supergravity dual to ${\cal N}=6$ Chern-Simons matter theory \cite{Aharony:2008ug}. This is the smallest space with the desired features, and given that a string vacua has to live in d=10 or 11, it does not take long to realise that the only additional option is to fiber a U(1) over $\mathbb{CP}^3$. Such solutions were ruled out in type II supergravity at the level of the equations of motion in \cite{Legramandi:2020txf} and those that exist in M-theory can always be reduced to IIA\footnote{Either the spinors are not charged under the additional U(1), or some algebra other than $\mathfrak{osp}(6|2)$ is being realised}. As such we seek to classify solutions of type II supergravity that are foliations of AdS$_3\times \mathbb{CP}^3$ over an interval, we leave the status of  d=11 vacua containing similar foliations over Riemann surfaces to be resolved elsewhere. 

For the algebra $\mathfrak{osp}(5|2)$ the 4-sphere is as much of a non-starter to realise the $\mathfrak{so}(5)$ R-symmetry as the 5-sphere was previously. One way to realise this algebra is to start with an existing $\mathfrak{osp}(6|2)$ solution, then orbifold $\mathbb{CP}^3$ by one of the discrete groups $\hat{D}_k$ (the binary dihedral group) or  $\mathbb{Z}_2$ , as discussed in the context of AdS$_4\times \mathbb{CP}^3$ in \cite{Aharony:2008gk}\footnote{see section 3 therein}. This however only breaks supersymmetry globally, locally such solutions still preserve $\mathfrak{osp}(6|2)$. One way, perhaps the only way\footnote{One can realise an $\mathfrak{so}(5)$ R-symmetry on a squashing of $\text{S}^3\rightarrow \text{S}^7\rightarrow\text{S}^4$, but AdS$_3$  vacua containing this factor only exists in $d=11$ and, when they support $\mathfrak{osp}(5|2)$, they can always be reduced to IIA within the 7-sphere resulting in squashed $\mathbb{CP}^3$  ($ \widehat{\mathbb{CP}}\!\!~^3$) and preserving ${\cal N}=(5,0)$.} to break to $\mathfrak{osp}(5|2)$ locally is to proceed  as follows: If one expresses $\mathbb{CP}^3$  as a fibration of S$^2$ over S$^4$ and then pinches the fiber, one breaks the SO(6) isometry down to SO(5) locally. The \textbf{6} of SO(6) branches as $\textbf{5}\oplus \textbf{1}$ under its SO(5) subgroup thereby furnishing us with both the representation and R-symmetry that $\mathfrak{osp}(5|2)$ demands. We shall thus also classify ${\cal N}=(5,0)$ AdS$_3$ vacua of type II supergravity on squashed $\mathbb{CP}^3$  by generalising the previous ansatz to included additional warp factors and SO(5) invariant terms in the flux.  We shall in fact classify both these vacua and those supporting $\mathfrak{osp}(6|2)$, or orbifolds there of,  in tandem as the latter are special cases of the former. 

We find two classes of solutions preserving respectively $\mathfrak{osp}(5|2)$ (locally) and $\mathfrak{osp}(6|2)$ superconformal algebras. We also find for each case that it is possible to construct solutions with bounded internal spaces, which should provide good dual descriptions of CFTs through the AdS/CFT correspondence.  The existence of backgrounds manifestly realising exactly the superconformal algebra $\mathfrak{osp}(5|2)$ is interesting in the light of \cite{Lee:2019uen}, which claims that all CFTs supporting such global superconformal algebras experience an enhancement to $\mathfrak{osp}(6|2)$. Our results cast some doubt on the veracity of  the claims of \cite{Lee:2019uen}, at least naively. It would be interesting to explore what leads to this apparent contradiction and whether this can be resolved, that however lies outside the scope of this work.\\
~\\
The layout of this paper is as follows:\\
~~\\
In section \ref{sec:bonsonicansatz} we consider AdS$_3$ vacua of type II supergravity that preserve an SO(5) isometry in terms of squashed $\mathbb{CP}^3$, without making reference to supersymmetry. On symmetry grounds alone we are able to give the local form that the NS and RR fluxes must take in regular regions of their internal space, which we found useful when deriving the results in the subsequent sections.

In section \ref{sec:nessisarysusy} we explain our method for solving the supersymmetry constraints. We reduce the problem to solving for a single ${\cal N}=1$ sub-sector of the full $(5,0)$ as the remaining 4 ${\cal N}=1$ sub-sectors  are shown to be implied by this and the action of $\mathfrak{osp}(5|2)$ which the spinors transform in. This enables us to employ an existing minimally supersymmetric AdS$_3$ bi-spinor classification \cite{Dibitetto:2018ftj,Passias:2019rga,Macpherson:2021lbr} to the case at hand.

In section \ref{ses:classification} we classify ${\cal N}=(5,0)$ vacua of type II supergravities realising the algebra $\mathfrak{osp}(5|2)$ in terms of a foliation of AdS$_3$   solutions of type II supergravity that are foliations of AdS$_3\times \widehat{\mathbb{CP}}\!\!~^3 $ over an interval, we are actually able to find the local form of all of them. They only exist in type IIA, generically have all possible fluxes turned on and are governed by two ODEs. The first of these takes the form $h'''=-2\pi F_0$, where $F_0$ is the Romans mass, making $h$ locally an order 3 linear polynomial highly reminiscent of the AdS$_7$ vacua of \cite{Apruzzi:2013yva,Cremonesi:2015bld}. The second ODE defines a linear function $u$ which essentially controls the squashing of $\mathbb{CP}^3$ and hence the breaking of  $\mathfrak{osp}(6|2)$ to $\mathfrak{osp}(5|2)$ . For generic values of $u$ one has ${\cal N}=(5,0)$ supersymmetry, but if one fixes $u=$ constant this is enhanced to ${\cal N}=(6,0)$

In section \ref{sec:local} we perform a regularity analysis of the local vacua establishing exactly what boundary behaviour are possible for the interval. We focus on ${\cal N}=(6,0)$ in section \ref{sec:osp62} where we find that fixing $F_0=0$ always gives rise to AdS$_4\times \mathbb{CP}^3$ locally, while for $F_0\neq 0$ it is possible to bound the interval at one end with several physical singularities but that the other end is always at infinite proper distance, at least when  $F_0$ is fixed globally. We study the ${\cal N}=(5,0)$ case in section \ref{sec:osp52} were conversely we find no AdS$_4$ limit and that a globally constant $F_0$ is no barrier to constructing bounded solutions. Many more physical boundary behaviours are possible in this case.

Up to this point in the paper we have assumed $F_0$ is constant, globally it need only be so piece-wise which allows for D8 branes along the interior of the interval - we explore this possibility in section \ref{sec:global}. We establish under what conditions such interior D8s are supersymmetric and explain how they can be used to construct broad classes of globally bonded solutions. We illustrate the point with some explicit examples. All of this  points the way to broad classes of duals interesting superconformal quiver we shall report on in \cite{USandYolanda}.

The work is supplemented by several appendices. In appendix \ref{sec:app1} we provide technical details of the construction of spinors on the internal space transforming in the fundamental representation of SO(5) and SO(6).  In appendix \ref{sec:bilinears} we present details of the $d=6$ bi-linears that feature during computations in section \ref{ses:classification}. Finally in appendix \ref{app:sop72nogo} we additionally show that all $\mathfrak{osp}(7|2)$ preserving AdS$_3$ vacua experience a local enhancement to AdS$_4\times$S$^7$ --  SO(7) preserving orbifolds of this are a possibility, but such constructions are AdS$_4$ rather than AdS$_3$ vacua.

\section{SO(2,2)$\times$SO(5) invariant type II supergravity on $\text{AdS}_3\times\widehat{\mathbb{CP}}\!\!~^3$ }\label{sec:bonsonicansatz}
In this section we consider the most possible vacua of type II supergravity that preserve the full SO(2,2)$\times$SO(5)  isometries of a warped product containing $\text{AdS}_3$ and a squashed $\mathbb{CP}^3$ ($\widehat{\mathbb{CP}}\!\!~^3$). Specifically we construct the full set of SO(5) invariant forms and use them to find the general form of NS and RR fluxes that are consistent with their source free (magnetic) Bianchi identities. Let us stress that this section makes no use of supersymmetry only symmetry, it is none the less useful when we choose to impose the former in the following sections.\\
~\\
In general AdS$_3$ solutions of type II supergravity admit a decomposition in the following form
\beq
ds^2= e^{2A} ds^2(\text{AdS}_3)+ ds^2(\text{M}_7),~~~H= e^{3A}h_0\text{vol}(\text{AdS}_3)+H_3,~~~ F=f_{\pm}+e^{3A}\text{vol}(\text{AdS}_3)\star_7\lambda(f_{\pm}), \label{eq:AdS3vacua}
\eeq
where $(e^{2A},f,H_3)$ and the dilaton $\Phi$ have support on M$_7$ so as to preserve the SO(2,2) symmetry of AdS$_3$. The $d=10$ NS and RR fluxes are $H$ and $F$ respectively, the latter expressed as a polyform of even/odd degree in IIA/IIB. The function $\lambda$ acts on a p-form as $\lambda(X_p)= (-1)^{[\frac{p}{2}]}X_p$ - this ensures the self duality constraint $F=\star_{10}\lambda (F)$.

We are interested in solutions where M$_7$ preserve an additional SO(5) isometry that can be identified with the R-symmetry of the ${\cal N}=(5,0)$ superconformal algebra $\mathfrak{osp}(5|2)$. The 4-sphere comes to mind as the obvious space realising an SO(5) isometry, however this supports Killing spinors in the \textbf{4} of SP(2), where as we require spinor in the \textbf{5} of SO(5) - so we will need to be more inventive.\\
~~\\ 
The coset space $\mathbb{CP}^3$ is a 6 dimensional compact manifold that can be generated by dimensionally reducing S$^7$ on its Hopf fiber - it appears most famously in the ${\cal N}=6$ AdS$_4\times \mathbb{CP}^3$ solution dual to Chern-Simons matter theory. The 7-sphere supports spinors transforming in the \textbf{8} of SO(8) and the reduction to  $\mathbb{CP}^3$ preserves the portion of these preserving the \textbf{6} of SO(6). Advantageously $\mathbb{CP}^3$  has a parametrisation as an S$^2$ fibration over S$^4$ that allows a squashing breaking SO(6)$\to$SO(5) by pinching the fiber - we will refer to this space as $\widehat{\mathbb{CP}}\!\!~^3$. As the \textbf{6} branches as $\textbf{1}\oplus \textbf{5}$ under SO(5) $\subset$ SO(6) clearly $\widehat{\mathbb{CP}}\!\!~^3$ supports both the isometry group and spinors we seek.  Embedding this SO(5) invariant space into M$_7$ leads to a metric ansatz of the form
\begin{align}
ds^2(\text{M}_7)&= e^{2k}dr^2+ ds^2(\widehat{\mathbb{CP}}\!\!~^3)\label{eq:defCP3squahsed}\\[2mm]
ds^2(\widehat{\mathbb{CP}}\!\!~^3)&= \frac{1}{4}\bigg[e^{2C} \left(d\alpha^2+ \frac{1}{4}\sin^2\alpha (L_i)^2\right)+ e^{2D} (Dy_i)^2\bigg],~~~~~ Dy_i= dy_i+\cos^2\left(\frac{\alpha}{2}\right)\epsilon_{ijk} y_j L_k,\nn
\end{align}
where $y_i$ are embedding coordinates on the unit radius 2-sphere, $L_i$ are a set of SU(2) left invariant forms and $(e^{2A},e^{2C},e^{2D},e^{2k},\Phi)$ have support on $r$ only.\\
~~\\
To write an ansatz for the fluxes on this space we need to construct the SO(5) invariant forms on $\widehat{\mathbb{CP}}\!\!~^3$. As explained in appendix \ref{sec:app1}, the S$^4$ base of this fiber bundle contains an SO(4)$=$SO(3)$_L\times$SO(3)$_R$ isometry in the 3-sphere spanned by $L_i$. In the full space SO(3)$_R$ is lifted to the diagonal SO(3) formed of  SO(3)$_R$ and the SO(3) of the 2-sphere. As such the invariants of SO(5) can be expanded in a basis of the SO(3)$_L\times$SO(3)$_D$ invariants on the S$^2\times $S$^3$ fibration (see for instance \cite{DeLuca:2018buk}), namely
\begin{align}
\omega_1&= \frac{1}{2}L_i y_i,~~~~\omega_2^1=\frac{1}{2} \epsilon_{ijk}y_i Dy_j\wedge D y_k,~~~~\omega_2^2=\frac{1}{2}L_i\wedge Dy_i,\nn\\[2mm]
\omega_2^3&=\frac{1}{2} \epsilon_{ijk}y_i L_j\wedge Dy_k,~~~~~\omega_2^4=\frac{1}{8}\epsilon_{ijk}y_i L_j\wedge L_k,\label{eq:SO3Dinvariants}
\end{align}
and wedge products there off,  leaving only the $\alpha$ dependence of the SO(5) invariants to fix via consistency with the remaining SO(5)/(SO(3)$_L\times$SO(3)$_D$) subgroup.

First off when $e^{2C}=e^{2D}=1$ we regain unit radius round $\mathbb{CP}^3$, which is a K\"ahler Einstein manifold with an SO(6) invariant K\"ahler form $J_2$, so we have the following SO(6) invariants on $\mathbb{CP}^3$
\beq
\mathbb{CP}^3:~e^{2C}=e^{2D}=1~~~\Rightarrow~~~ \text{SO(6) invariants}:~~ J_2,~~~ J_2\wedge J_2,~~~J_2\wedge J_2\wedge J_2= 6 \text{vol}(\mathbb{CP}^3).
\eeq
where specifically
\beq
J_2= \frac{1}{4}\bigg(\sin\alpha d\alpha \wedge \omega_1- \sin^2\alpha \omega_2^4- \omega_2^1\bigg)\label{eq:J}.
\eeq
It is not hard to show that the remaining SO(5) invariants, which are not invariant under the full SO(6) of $\mathbb{CP}^3$, may be expressed in terms of the SU(3)-structure spanned by
\beq
\tilde J_2= \frac{1}{4}\bigg(\sin\alpha d\alpha \wedge \omega_1- \sin^2\alpha \omega_2^4+ \omega_2^1\bigg),~~~~\Omega_3=-\frac{1}{8}\sin\alpha\bigg(\sin\alpha \omega_1+i d\alpha\bigg)\wedge \bigg(\omega^3_2+i\omega_2^2\bigg)\label{eq:SO5s},
\eeq
These invariant forms obey the following identities
\begin{align}
&J_2\wedge \Omega_3=\tilde{J}_2\wedge \Omega_3=0,~~~~J_2\wedge J_2\wedge J_2=\tilde J_2\wedge \tilde J_2\wedge \tilde J_2=\frac{3 i}{4}\Omega_3\wedge\overline{\Omega}_3,~~~~J_2\wedge J_2+ \tilde J_2\wedge \tilde J_2=2 \tilde J_2\wedge J_2,\nn\\[2mm]
&d J_2=0,~~~~ d\tilde J_2= 4 \text{Re}\Omega_3,~~~~d \text{Im}\Omega_3=6 \tilde J_2\wedge J_2-2 J_2\wedge J_2,
\end{align}
and as such, they form a closed set under the exterior product and derivative. This is all that is needed to construct the fluxes.\\
~\\
The general form of an SO(5) invariant $H_3$ obeying $dH_3=0$ is given by
\beq
H_3= dB_2,~~~~B_2=  b(r)J_2+\tilde{b}(r) \tilde {J}_2,\label{eq:magneticNSflux}
\eeq
The general SO(5) invariant $f_{\pm}$ obeying $df_{\pm}=H_3\wedge f_{\pm}$ can be expressed as
\begin{align}
f_+&= \bigg[F_0+c_1 J_2+c_2 J_2\wedge J_2+c_3\frac{1}{3!}J_2\wedge J_2\wedge J_2+ d\bigg(p(r)\text{Im}\Omega_3+ q(r)\text{Re}\Omega_3\bigg)\bigg]\wedge e^{B_2},\nn\\[2mm]
f_-&= d\bigg[a_1(r)+a_2(r) J_2+a_3(r)\frac{1}{2}J_2\wedge J_2+a_4(r)J_2\wedge \tilde{J}_2+a_5(r)\frac{1}{3!}J_2\wedge J_2\wedge J_2\bigg]\wedge e^{B_2},
\end{align}
giving us an SO(5) invariant ansatz for the flux in IIA/IIB which is valid away from the loci of localised sources\footnote{These need to be generalised in scenarios which allow for sources smeared over all their co-dimensions}. In IIA this depends locally on 4 constants $(F_0,c_1,c_2,c_3)$ and 4 functions of $r $ $(b,\tilde {b},p,q)$ - there is an enhancement to SO(6) when $\tilde{b}=p=q=0$.  If we also consider $d(e^{3A}\star_7f_{\pm})=e^{3A}H_3\wedge \star_7f_{\pm}$ we find we must in general fix $q=0$. In IIB this depends on 7 functions of $r$, with an enhancement to SO(6) when $\tilde{b}=a_4=0$.

\section{Necessary and sufficient conditions for realising supersymmetry}\label{sec:nessisarysusy}
In this section we present the method by which we shall impose supersymmetry on SO(5) invariant ansatz of the previous section.\\
~\\
Geometric conditions for ${\cal N}=(1,0)$ AdS$_3$ solutions with purely magnetic NS flux (ie $h_0=0$) were derived first in massive IIA in \cite{Dibitetto:2018ftj}, then generalised to IIB in \cite{Passias:2019rga} with the assumption that $h_0=0$, this assumption was then relaxed in \cite{Macpherson:2021lbr}  whose conventions we shall follow. These conditions are defined in terms of two non vanishing Majorana spinors $(\hat{\chi}_1,\hat{\chi}_2)$ on the internal M$_7$ which without loss of generality obey
\beq
|\hat{\chi}_1|^2+ |\hat{\chi}_2|^2= 2 e^{A},~~~~ |\hat{\chi}_1|^2- |\hat{\chi}_2|^2= c e^{-A},
\eeq
for $c$ an arbitrary constant. One can solve these constraints in general in terms of two unit norm spinors $(\chi_1,\chi_2)$ and a point dependent  angle $\theta$ as
\beq
\hat{\chi}_1=e^{\frac{A}{2}}\sqrt{1-\sin\theta}\chi_1,~~~~ \hat{\chi}_2=e^{\frac{A}{2}}\sqrt{1+\sin\theta}\chi_2,~~~ c=-2e^{2A} \sin\theta.
\eeq
Plugging this into the necessary and sufficient conditions for supersymmetry in \cite{Macpherson:2021lbr} (see Appendix B therein), we find they become\footnote{These do not represent a set of necessary and sufficient conditions when $\cos\theta=0$. However as this limit turns off one of $(\hat{\chi}_1,~\hat{\chi}_2) $ the NS 3-form is the only flux that can be non trivial. The common NS sector of type II supergravity is S-dual to classes of IIB solution with the RR 3-form the only non trivial flux which are contained in the conditions we quote. }
\begin{subequations}
\begin{align}
&e^{3A}h_0=  2 m e^{2A}\sin\theta,~~~~d(e^{2A}\sin\theta)=0,\label{eq:SUSY0}\\[2mm]
&d_{H_3}(e^{2A-\Phi}\cos\theta\Psi_{\mp})= \pm \frac{1}{8}e^{2A}\sin\theta f_{\pm},\label{eq:SUSYI}\\[2mm]
&d_{H_3}(e^{3A-\Phi}\cos\theta\Psi_{\pm})\mp 2m e^{2A-\Phi}\cos\theta\Psi_{\mp}=\frac{e^{3A}}{8}\star_7\lambda(f_{\pm}),\label{eq:SUSYII}\\[2mm]
&e^{A}(\Psi_{\mp},~f_{\pm})_7=\mp\frac{m}{2}e^{-\Phi}\cos\theta \text{vol}(\text{M}_7),\label{eq:projection}
\end{align}
\end{subequations}
where $(\Psi_{\mp},~f_{\pm})_7$ is the 7-form part of $\Psi_{\mp}\wedge \lambda (f_{\pm})$ and the real even/odd bi-linears $\Psi_{\pm}$ are defined via
\beq
\chi_1\otimes \chi_2^{\dag}= \frac{1}{8}\sum_{n=0}^7\frac{1}{n!}\chi_2^{\dag}\gamma_{a_n...a_1}\chi_1e^{a_1...a_n}=\Psi_++i \Psi_-
\eeq
for $e^{a}$ a vielbein on M$_7$. In the above $m$ is the inverse AdS$_3$ radius, in particular when $m=0$ we have Mink$_3$ while when $m \neq 0$ its precise value is immaterial as it can be absorbed into the AdS$_3$ warp factor, thus going forward we fix
\beq
m=1
\eeq
without loss of generality.\\
~\\
In this work we will construct explicit solutions preserving  $(5,0)$ and $(6,0)$ supersymmetries and for the cases of extended supersymmetry \eqref{eq:SUSY0}-\eqref{eq:projection} is not on its own sufficient. If one has ${\cal N}=(n,0)$ supersymmetry one has $n$ independent ${\cal N}=(1,0)$ sub-sectors that necessarily come with their corresponding $n$ independent bi-linears $\Psi^{(n)}_{\pm}$. These must all solve \eqref{eq:SUSY0}-\eqref{eq:projection} for the same bosonic fields of supergravity. However the AdS$_3$ vacua we are interested in realise the superconformal algebra $\mathfrak{osp}(n|2)$ which means the internal spinors which define these bi-linears transform in the $\textbf{n}$ of $\mathfrak{so}(n)$ while the bosonic fields are $\mathfrak{so}(n)$ singlets. Thus the bi-linears decompose into parts transforming in irreducible representations of the tensor product $\textbf{n}\otimes \textbf{n}$. Specifically this contains a singlet part that is common to all $\Psi^{(n)}_{\pm}$  and a charged part in the symmetric representation\footnote{Of course $\textbf{n}\otimes \textbf{n}$ decomposes into singlet, symmetric traceless and anti-symmetric representations, however to see the anti-symmetric representation one would need to construct bi-linears that mix the internal spinors $\chi_{1}$ and $\chi_{2}$ that belong to different $(1,0)$ sub-sectors - this is not needed for our purposes.}. The charged parts of $\Psi^{(n)}_{\pm}$ are mapped into each other  by taking the Lie derivative with respect to the SO(n) Killing vectors, and in particular the bi-linears of a single $(1,0)$ sub-sector + the action of SO(n) is enough to generate the whole set. Then, since the Lie and exterior derivatives commute, it follows that if a single pair of bi-linears, $\Psi^{1}_{\pm}$ say, solve \eqref{eq:SUSY0}-\eqref{eq:projection} then they all do. 

In summary to know that extended supersymmetry holds on \eqref{eq:defCP3squahsed} it is sufficient to construct an n-tuplet of spinors that transform in the $\textbf{n}$ of $\mathfrak{so}(n)$, and then solve \eqref{eq:SUSY0}-\eqref{eq:projection} for the $\Psi_{\pm}$ following from any ${\cal N}=(1,0)$ sub-sector whilst imposing that the bosonic fields are  all $\mathfrak{so}(n)$ singlets. In particular this means that we must solve \eqref{eq:SUSY0}-\eqref{eq:projection} under the assumption that the warp factors and dilaton only depend on $r$ and the fluxes only depend on $r$ and SO(n) invariant forms. We deal with the bulk of the construction of these SO(n) spinors in appendix \ref{sec:app1} where we construct spinors in relevant representations on $\mathbb{CP}^3$.  Below we present the embedding of these spinors into \eqref{eq:defCP3squahsed}. \\
~\\
${\cal N}=6$ spinors in $d=7$ can be expressed in terms of 4 real functions of $r$ $(f_1,f_2,g_1,g_2)$ and the spinors in the \textbf{6} of SO(6) on $\mathbb{CP}^3$ in \eqref{eq:the6}
\begin{align}
\chi^{{\cal I}}_1&= \cos\left(\frac{\beta_1+\beta_2}{2}\right) \xi^{{\cal I}}_6+  i  \sin\left(\frac{\beta_1+\beta_2}{2}\right)\gamma_7\xi^{{\cal I}}_6,\nn\\[2mm]
\chi^{{\cal I}}_2&= \cos\left(\frac{\beta_1-\beta_2}{2}\right) \xi^{{\cal I}}_6+  i \sin\left(\frac{\beta_1-\beta_2}{2}\right)\gamma_7\xi^{{\cal I}}_6,\label{eq:neq6spinors}
\end{align}
where ${\cal I}=1,...,6$ and $\beta_{1,2}=\beta_{1,2}(r)$. These are only valid on round $\mathbb{CP}^3$, ie when $e^{2B}=e^{2C}$ and the fluxes depend on $\mathbb{CP}^3$ through the SO(6) invariant 2-form $J_2$. We will not actually make explicit use of these spinors as it turns out that general class of ${\cal N}=(6,0)$ is actually simply one of 2 branching classes of solution following from the ${\cal N}= (5,0)$ spinors below.\\
~\\
${\cal N}=5$ spinors in $d=7$ can be decomposed in terms the spinors in the \textbf{5} of SO(5) on $\mathbb{CP}^3$  in \eqref{eq:spinorreps} and 4 constraints as
\begin{align}
\chi^{\alpha}_1&= a_{11}(\xi_5^{\alpha}+Y_{\alpha} i \gamma_7\xi_0)+b_{11}(i\gamma_7\xi_5^{\alpha}-Y_{\alpha} \xi_0)+a_{12} Y_{\alpha}\xi_0+i b_{12} Y_{\alpha}\gamma_7\xi_0,\nn\\[2mm]
\chi^{\alpha}_2&= a_{21}(\xi_5^{\alpha}+Y_{\alpha} i \gamma_7\xi_0)+b_{21}(i\gamma_7\xi_5^{\alpha}-Y_{\alpha}\xi_0)+a_{22} Y_{\alpha}\xi_0+i b_{22} Y_{\alpha}\gamma_7\xi_0,\nn\\[2mm]
& a_{11}^2+b_{11}^2=a_{12}^2+b_{12}^2= a_{21}^2+b_{21}^2=a_{22}^2+b_{22}^2=1\label{eq:norconstraints}.
\end{align}
where $\alpha=1,...,5$, the 8 parameters $a_{11},b_{11},...$ are all real and have support on $r$ alone,  we have parameterised things in this fashion to make the unit norm constraints simple. These spinors are valid for squashed $\mathbb{CP}^3$.\\
~\\
Finally a set of  ${\cal N}=1$  spinors can also be defined in $d=7$, they are given by 
\begin{align}
\chi^{(0)}_1&= \cos\left(\frac{\beta_1+\beta_2}{2}\right) \xi_0+  i  \sin\left(\frac{\beta_1+\beta_2}{2}\right)\gamma_7\xi_0,\nn\\[2mm]
\chi^{(0)}_2&= \cos\left(\frac{\beta_1-\beta_2}{2}\right) \xi_0+  i  \sin\left(\frac{\beta_1-\beta_2}{2}\right)\gamma_7\xi_0,
\end{align}
where again $\beta_{1,2}=\beta_{1,2}(r)$ and these spinors are valid on squashed $\mathbb{CP}^3$. The $0$ superscript refers to the the fact that these are SO(5) $\subset$ SO(6) singlets. These are in fact nothing more than the 6th component of \eqref{eq:neq6spinors}, however unlike the ${\cal N}=(6,0)$ case $ e^{2C}\neq  e^{2B}$ and the flux can depend on more than merely $r$ and $J_2$. These spinors can be used to construct ${\cal N}=(1,0)$ AdS$_3$ solutions with SO(5) flavour symmetry, something we will report on elsewhere  \cite{US}.

\section{Classification of $\mathfrak{osp}(n|2)$ AdS$_3$ vacua on $\widehat{\mathbb{CP}}\!\!~^3$ for $n=5,6$}\label{ses:classification}
In this section we classify AdS$_3$ solutions preserving ${\cal N}=(5,0)$ supersymmetry on squashed $\mathbb{CP}^3$. Such solutions only exist in type IIA supergravity and experience an enhancement to  ${\cal N}=(6,0)$ when a function is appropriately fixed. We summarise our results between \eqref{eq:summerystart} and \eqref{eq:summeryend}.\\
~\\
We take our representative ${\cal N}=1$ sub-sector to be
\beq
\chi_1=\chi^{5}_1,~~~~\chi_2= \chi^5_2,
\eeq
which has the advantage that the bi-linears decompose in terms of the SO(3)$_L\times$SO(3)$_D$ invariant forms on the S$^2\times$S$^3$ fibration.
We find the  $d=7$ bi-linears are given by
\beq\label{eq:so5bilinears}
\Psi_+= (\underline{{\cal S}}^1)_+. (\underline{\phi})_+ +e^k   (\underline{{\cal S}}^2)_- .(\underline{\phi})_- \wedge dr,~~~~\Psi_-=  (\underline{{\cal S}}^1)_-. (\underline{\phi})_-+ e^k  (\underline{{\cal S}}^2)_+ .(\underline{\phi})_+ \wedge dr
\eeq
where we define
\beq
(\underline{\phi})_{\pm}=(\phi^1_{\pm},~\phi^2_{\pm},~Y_5\phi^3_{\pm},~Y_5\phi^4_{\pm},~Y_5\phi^5_{\pm},~Y_5\phi^6_{\pm},~Y_5^2\phi^7_{\pm},~Y_5^2\phi^8_{\pm}),~~~~~Y_5=\cos\alpha,
\eeq
for $\phi^1_{\pm}$ real even/odd bi-linears on $\widehat{\mathbb{CP}}\!\!~^3$ decomposing in a basis of \eqref{eq:SO3Dinvariants} - their explicit form is given in \eqref{eq:SO3Dpolyforms}. We also define
\begin{align}
(\underline{{\cal S}}^1)_+&=\left(\begin{array}{c}a_{11}a_{21}+b_{11}b_{21}\\a_{11}b_{21}-a_{21}b_{11}\\a_{11}a_{22}+b_{11}b_{22}\\a_{11}b_{22}- a_{22}b_{11}\\a_{12}a_{21}+b_{12}b_{21}\\a_{21}b_{12}- a_{12}b_{21}\\a_{12}a_{22}+b_{12}b_{22}\\a_{22}b_{12}-a_{12}b_{22}\end{array}\right),~~~~~(\underline{{\cal S}}^2)_-=\left(\begin{array}{c}a_{21}b_{11}+a_{11}b_{21}\\b_{11}b_{22}-a_{11}a_{21}\\a_{11}b_{22}+a_{22}b_{11}\\b_{11}b_{22}-a_{11}a_{22}\\a_{21}b_{12}+a_{12}b_{21}\\b_{12}b_{21}-a_{12} a_{21}\\a_{22}b_{12}+a_{12}b_{22}\\b_{12}b_{22}-a_{12}a_{22}\end{array}\right),\nn\\[2mm]
(\underline{{\cal S}}^1)_-&=\left(\begin{array}{c}a_{21}b_{11}-a_{11}b_{21}\\a_{11}a_{21}+b_{11}b_{21}\\a_{22}b_{11}-a_{11}b_{22}\\a_{11}a_{22}+b_{11}b_{22}\\a_{21}b_{12}-a_{12}b_{21}\\-a_{12}a_{21}-b_{12}b_{21}\\a_{22}b_{12}-a_{12}b_{22}\\-a_{12}a_{22}-b_{12}b_{22}\end{array}\right),~~~~(\underline{{\cal S}}^2)_+=\left(\begin{array}{c}a_{11}a_{21}-b_{11}b_{21}\\a_{21}b_{11}+a_{11}b_{21}\\a_{11}a_{22}-b_{11}b_{22}\\a_{22}b_{11}+a_{11}b_{22}\\a_{12}a_{21}-b_{12}b_{21}\\a_{21}b_{12}+a_{12}b_{21}\\a_{12}a_{22}-b_{12}b_{22}\\a_{22}b_{12}+a_{12}b_{22}\end{array}\right).
\end{align}
We begin by solving the constraints in \eqref{eq:norconstraints} by parametrising the functions of the spinor ansatz as
\begin{align}
a_{11}+i b_{11}&=e^{\frac{i}{2}(X_1+X_3)},~~~~a_{12}+i b_{12}=e^{\frac{i}{2}(X_2+X_4)},\nn\\[2mm]
a_{21}+i b_{21}&=e^{\frac{i}{2}(X_1-X_3)},~~~~a_{22}+i b_{22}=e^{\frac{i}{2}(X_2-X_4)},
\end{align}
for $X_{1}...X_4$ functions of $r$ only. We shall take the magnetic component of the NS 3-form as in \eqref{eq:magneticNSflux} and allow the RR fluxes to depend on $r$ and all the SO(5) invariant forms, ie
\beq
dr,~~~J_2,~~~\tilde J_2,~~~ \text{Re}\Omega_3,~~~\text{Im}\Omega_3,
\eeq
and the wedge products one can form out of these. One then proceed to substitute \eqref{eq:so5bilinears} into the necessary conditions for supersymmetry \eqref{eq:SUSY0}-\eqref{eq:projection} to fix the $r$ dependence of the ansatz.\\ 
~~\\
In IIB supergravity there are no solutions: One arrives at a set of algebraic constraints by solving for the parts of \eqref{eq:SUSYI}-\eqref{eq:SUSYII} orthogonal to $dr$ which without loss of generality fix the phases as
\beq
X_1=-\frac{\pi}{2}+2\beta(r),~~~~X_2=\frac{\pi}{2},~~~~\theta=X_3=X_4=0.
\eeq
and several parts of the metric and NS 2-form as 
\beq
e^{C}=5 e^{A}\sin\beta,~~~~e^D= 5 e^A\sin\beta \cos\beta,~~~~\tilde{b}=0.
\eeq
Unfortunately if one then tries to solve the $dr$ dependent terms in \eqref{eq:SUSYI} one finds the constraint
\beq
\cos\beta=0,
\eeq
which cannot be solved without setting $e^D=0$, so no ${\cal N}=(5,0)$ or $(6,0)$ solutions exist on this space in type IIB.\\
~~\\
Moving onto type IIA supergravity: Some conditions one may extract from \eqref{eq:SUSYI}, which simplify matters considerably going forward, are the following
\beq
\sin\theta=0,~~~~\sin X_1=-\sin X_2=1,
\eeq
which we can solve without loss of generality as 
\beq
\theta=0~~~~\Rightarrow~~~~ h_0=0,~~~~X_1=-X_2= \frac{\pi}{2}.
\eeq
We then choose to further refine the phases  as
\beq
X_3=\beta_1+\beta_2,~~~~ X_4= -\beta_1+\beta_2.
\eeq
Plugging these into \eqref{eq:SUSYI}-\eqref{eq:projection} we find the following simple definitions of various functions in the ansatz
\begin{align}
 e^{C}&=2 e^{A}\sin \beta_2,~~~~ e^D= 2 e^{A} \sin(\beta_1+\beta_2),\nn\\[2mm]
b'&=4 e^{A+k}+2 \partial_{r}(e^{2A}\cos \beta_1 \sin(\beta_1+2 \beta_2)),~~~~ \tilde{b}= -2 e^{2A} \cos(\beta_1+2\beta_2)\sin \beta_1,\label{eq:defsofbosonicfileds}
\end{align}
and the following ODEs that need to be actively solved
\begin{align}
&\partial_r(e^{3A-\Phi}\sin\beta_1 \sin\beta_2)+2m e^{2A+k-\Phi}\sin\beta_1 \cos\beta_2=0,\nn\\[2mm]
&\partial_r(e^{5A-\Phi}\sin^2\beta_2 \sin(\beta_1+\beta_2))+m e^{4A+k-\Phi}\sin\beta_2\sin(\beta_1+2 \beta_2)=0,\nn\\[2mm]
&\partial_r(e^{2A}\tan\beta_2)+m e^{A+k}(2\tan\beta_2\cot(\beta_1+\beta_2)-(\cos\beta_2)^{-2})=0\label{eq:bpsneq53}.
\end{align}
We also extract expressions for the RR fluxes, though we delay presenting them  explicitly until we have simplified the above.
To make progress we find it useful to use diffeomorphism invariance in $r$ to fix\footnote{The reason for the factors of $\pi$, taken here and elsewhere without loss of generality, is that they make the Page charges of the RR fluxes simple.}
\beq
e^{A+k}=-\pi,
\eeq
 and introduce local functions of $r$  $(h,g,u)$ such that
\beq
e^{5A-\Phi}\sin^2\beta_2 \sin(\beta_1+\beta_2)=\pi^2h,~~~~e^{2A}\tan\beta_2= g,~~~~\frac{\tan\beta_2}{\tan(\beta_1+\beta_2)}= \frac{h'+ h \frac{u'}{u}}{h'- h \frac{u'}{u}}\label{eq:ghdefs}.
\eeq
This simplifies the system of ODEs in \eqref{eq:bpsneq53} to
\begin{align}
u''&=0,~~~~g= 2\pi  \frac{hu}{u h'-hu'},~~~~\tan\beta_1=\frac{\text{sign}(u h)u'\sqrt{\Delta_1}}{u'(u h'-h u')+u^2 h''},~~~~\tan\beta_2= \frac{\text{sign}(u h)\sqrt{\Delta_1}}{u h'- h u'}\nn\\[2mm]
\Delta_1&=2 h h'' u^2-(u h'- h u')^2\label{eq:simpeodes}.
\end{align}
which imply supersymmetry and require $\Delta_1>0$.  What remains is the explicit form of the magnetic components of the RR fluxes. These can be expressed most succinctly in terms of their Page flux avatars, ie $\hat f_+ = f_+\wedge  e^{-B_2}$, however to compute these we must first integrate $b'$. Combining \eqref{eq:ghdefs}, \eqref{eq:simpeodes} and \eqref{eq:defsofbosonicfileds} we find
\beq
b=-\tilde{b}+4\pi \left(-(r-k)+\frac{u h'- h u'}{u h''}\right),~~~~~\tilde b=- 2\pi \frac{ u'}{h''}\left(\frac{h}{u }+ \frac{h h''-2 (h')^2}{2h' u'+u h''}\right),
\eeq
where $k$ is an integration constant. We then find for the magnetic Page fluxes 
\begin{align}
\hat f_0&=F_0= -\frac{1}{2\pi}h''',\nn\\[2mm]
\hat f_2&= 2 (h''-(r-k) h''')J,\nn\\[2mm]
\hat f_4&= -4\pi \bigg[(2h'+(r-k)(-2 h''+(r-k) h'''))J_2\wedge J_2+ d\left(\frac{h u'}{u}\text{Im}\Omega_3\right)\bigg],\nn\\[2mm]
\hat f_6&= \frac{16 \pi^2}{3}(6h-(r-k)(6h'+(r-k)(-3 h''+(r-k) h''')))J_2\wedge J_2\wedge J_2,\label{eq:pagefluxess}
\end{align}
where we have made extensive use of the conditions derived earlier to simply these expressions. In order to have a solution we must impose that Bianchi identities of the RR flux hold (that of the NS 3-form is implied), away from sources this is equivalent to imposing that $d\hat f_{2n}=0$ for $n=0,1,2,3$, we find
\beq
 d \hat f_{2n}=-\frac{1}{2\pi}(4\pi)^n\frac{1}{n!}(r-k)^n h'''' dr\wedge J_2^n,\label{eq:bianchis}
\eeq
which tells us that the Bianchi identities in regular parts of the internal space demand
\beq
h''''=0,
\eeq
or in other words that $h$ is an order 3 polynomial (at least locally). This completes our local derivation of the class of solutions.\\
~~\\
In summary the local form of solutions in this class take the following form: NS sector
\begin{align}
\frac{ds^2}{2\pi}&= \frac{|hu|}{\sqrt{\Delta_1}}ds^2(\text{AdS}_3)+\frac{\sqrt{\Delta_1}}{4|u|}\bigg[ \frac{2}{ |h''|}\bigg( ds^2(\text{S}^4)+ \frac{1}{ \Delta_2}(Dy_i)^2\bigg)+\frac{1}{|h|}dr^2\bigg],\nn\\[2mm]
e^{-\Phi}&=\frac{ \sqrt{|u|}|h''|^{\frac{3}{2}}\sqrt{\Delta_2}}{2\sqrt{\pi}\Delta_1^{\frac{1}{4}}},~~~~\Delta_1=2 h h'' u^2-(u h'- h u')^2,~~~~\Delta_2=1+ \frac{2 h' u'}{u h''},~~~~H=dB_2,\nn\\[2mm]
 B_2&= 4\pi \bigg[\left(-(r-k)+\frac{u h'-h u'}{u h''}\right) J_2+ \frac{u'}{2h''}\left(\frac{h}{u}+\frac{h h''-2 (h')^2}{2h'u'+uh''}\right)\left(J_2- \tilde J_2\right)\bigg],\label{eq:summerystart}
\end{align}
where $(u,h)$ are functions of $r$ and $k$ is a constant. Note that positivity of the metric and dilaton holds whenever $\Delta_1\geq 0$\footnote{Specifically reality of the metric demands $\Delta_1\geq 0$ and that $(h, u)$ are real. This in turn implies $h h''\geq 0$ and it then follows that $\Delta_2\geq 0$. Note that one needs to use that $h h''\geq 0$ to bring the metric to this form.}.  The $d=10$ RR fluxes are given by
\begin{align}
F_0&=-\frac{1}{2\pi}h''',~~~~~F_2= B_2 F_0+ 2(h''-(r-k) h''')J_2,\nn\\[2mm]
F_4&=\pi d\left(h'+\frac{hh''u(uh'+ h u')}{\Delta_1}\right)\wedge \text{vol}(\text{AdS}_3)+B_2\wedge F_2-\frac{1}{2}B_2\wedge B_2 F_0\nn\\[2mm]
&- 4\pi \bigg[(2h'+(r-k)(-2 h''+(r-k) h'''))J_2\wedge J_2+ d\left(\frac{h u'}{u}\text{Im}\Omega_3\right)\bigg].
\end{align}
Solutions within this class are defined locally by 2 ODEs:
First supersymmetry demands 
\beq
u''=0,\label{eq:bpscondtion}
\eeq
which must hold globally. Second the Bianchi identities of the fluxes demand that in regular regions of the internal space
\beq
h''''=0,\label{eq:summeryend}
\eeq
which one can integrate as 
\beq
h= c_0+ c_1 r+ \frac{1}{2}c_2 r^2+\frac{1}{3!}c_3 r^3,
\eeq
where $c_i$ are integration constants here and elsewhere. However the RHS of \eqref{eq:summeryend} can contain $\delta$-function sources globally, as we shall explore in section \ref{sec:global}. Before moving on to analyse solutions within this class it is important to stress a few things. First off that \eqref{eq:bpscondtion} must hold globally, and given how $u,u'$ appear in the class means that really $u$ is parametrising two branching possibilities - either $u'=0$ or $u'\neq 0$.

For the first case notice that if $u'=0$ then $u$ actually completely drops out of the bosonic fields, so its precise value  doesn't matter. Further the warping of the 4-sphere and fibered 2-sphere becomes equal, making the metric on $\mathbb{CP}^3$ the round one, and  only $J_2$ now appears in the fluxes. There is thus an enhancement of the global symmetry of the internal space to SO(6) - indeed supersymmetry is likewise enhanced to ${\cal N}=(6,0)$. We shall study this limit in section \ref{sec:osp62}.

When $u'\neq 0$ then  $u$ is an order 1 polynomial, however the class is invariant under $(r\to r+  l,~ k\to k+l)$ which one can use to set the constant term in $u$ to zero without loss of generality, the specific value of the constant $u'$ then also drops out of the bosonic fields. Thus for the second class, preserving only ${\cal N}=(5,0)$ supersymmetry, one can fix $u=r$ without loss of generality. We shall study this limit in section \ref{sec:osp52}.

Having a classes of solutions defined in terms of the ODE  $h'''= -2\pi  F_0$ is very reminiscent of AdS$_7$ vacua in massive IIA, which obey essentially the same constraint \cite{Cremonesi:2015bld}. For the $(6,0)$ case the formal similarities become more striking as both this and the AdS$_7$ vacua are of the form AdS$_{2p+1}\times\mathbb{CP}^{4-p}$ foliated over an interval in terms of an order 3 polynomial and it's derivatives - however we should stress these functions do not appear in the same way in each case. None the less this apparent series of local solutions does beg the question, what about  AdS$_5\times \mathbb{CP}^{2}$?. Establishing whether this also exists, and how much if any supersymmetry it may preserve, is beyond the scope of this work but would be interesting to pursue.

\section{Local analysis of $\mathfrak{osp}(n|2)$ vacua for $n>4$}\label{sec:local}
In this section we perform a local analysis of the  $\mathfrak{osp}(6|2)$ and $\mathfrak{osp}(5|2)$ AdS$_3$ vacua derived in the previous sections in  \ref{sec:osp62} and \ref{sec:osp52}. We begin with some comments about the non existence of  $\mathfrak{osp}(n|2)$ AdS$_3$ for $n>6$ and on the generality of classes we do find. \\
\\
In the previous section we derived classes of solutions in IIA supergravity that realise the super conformal algebras $\mathfrak{osp}(n|2)$ for $n=5,6$ on a warped product space consisting of a foliation of  AdS$_3\times  \widehat{\mathbb{CP}}\!\!~^3$ foliated over an interval such that either SO(5) or SO(6) is preserved.  In appendix \ref{app:sop72nogo} we prove that the case of $n=7$ is locally AdS$_4\times$S$^7$, and the same is proved for the case of $n=8$ in \cite{Legramandi:2020txf}. Thus one only has true AdS$_3$ solutions for $n=5,6$ (or lower) and we have found them only in type IIA. 

The class of $\mathfrak{osp}(6|2)$ AdS$_3$ vacua we find are exhaustive for type II supergravities: Spinors transforming in the \textbf{6} of $\mathfrak{so}(6)$ necessitate either a round $\mathbb{CP}^3$ factor in the metric or $\mathbb{CP}^3$ with a U(1) fibered over it. This latter possibility can easily be excluded at the level of the equations of motion \cite{Legramandi:2020txf}. For the case of $\mathfrak{osp}(5|2)$ AdS$_3$ vacua we suspect the same is true, but are not completely certain of that.

\subsection{ Analysis of $\mathfrak{osp}(6|2)$ vacua}  \label{sec:osp62}
The $\mathfrak{osp}(6|2)$ AdS$_3$ solutions are given by the class of the previous section specialised to the case $u=1$. We find for the NS sector
\begin{align}
\frac{ds^2}{2\pi}&= \frac{|h|}{\sqrt{2 h h'' -( h')^2}}ds^2(\text{AdS}_3)+\sqrt{2 h h'' -( h')^2}\bigg[\frac{1}{4|h| }dr^2+ \frac{2}{ |h''|} ds^2(\mathbb{CP}^3)\bigg],\nn\\[2mm]
e^{-\Phi}&=\frac{ (|h''|)^{\frac{3}{2}}}{2\sqrt{\pi}(2 h h'' -( h')^2)^{\frac{1}{4}}},~~~~
H=dB_2,~~~~ B_2= 4\pi\left(-(r-k)+\frac{ h'}{ h''}\right) J_2,\label{eq:neq6nssector}
\end{align}
and the RR sector
\begin{align}
F_0&=-\frac{1}{2\pi}h''',~~~~~F_2= B_2 F_0+  2(h''-(r-k) h''')J_2,\nn\\[2mm]
F_4&=\pi d\left(h'+\frac{hh'h''}{2 h h'' -( h')^2}\right)\wedge \text{vol}(\text{AdS}_3)+B_2\wedge F_2-\frac{1}{2}B_2\wedge B_2 F_0\nn\\[2mm]
& -4\pi(2h'+(r-k)(-2 h''+(r-k) h'''))J_2\wedge J_2,
\end{align}
where $k$ is a constant and $h$ is a function of $r$ obeying $h''''=0$ in regular regions of a solution.
\subsubsection{Local solutions and regularity }
There are several distinct  physical behaviours one can realise locally by solving  $h'''=-2\pi F_0$  (for $F_0=$ constant) in different ways, in this section we shall explore them.\\
~\\
The distinct local solutions the class contains can be characterised as follows. First off the domain of $r$ should be ascertained, in principle it can be one of the following: Periodic, bounded  (from above and below), semi infinite or unbounded. For a well defined AdS$_3$ vacua dual to a $d=2$ CFT $r$ must be one of the first 2. In the case at hand the $r$ dependence of $h$ does not allow for periodic $r$ so we seek bounded solutions. In general a  solution can be bounded by either a regular zero or a physical singularity.

At a regular zero we must have that $e^{-\Phi}$ and the AdS warp factor becomes constant. The internal space should then decompose as a direct product of  two sub manifolds with the first tending to the behaviour of a Ricci flat cone of radius $r$ and the second $r$ independent.

There are many ways to realise physical singularities that bound the space at some loci. The most simple is with D branes and O-planes: For a generic solution these objects are characterised by a metric and dilaton which decompose as
\beq
ds^2= \frac{1}{\sqrt{h_{p}}}ds^2_{\|}+ \sqrt{h_{p}} ds^2_{\perp},~~~~e^{-\Phi}\propto h_{p}^{\frac{p-3}{4}}
\eeq
where $ds^2_{\|}$ is the $p+1$ dimensional metric on the world volume of this object and $ds^2_{\perp}$ is the  $9-p$ dimensional metric on its co-dimensions.  We will consider only solutions whose metric is a foliation over an interval $r$. A Dp  brane singularity (for $p<7$) is then signaled by the leading order behaviour
\beq
ds^2_{\perp}\propto dr^2+ r^2 ds^2(\text{B}^{8-p}),~~~ h_p\propto \frac{1}{r^{7-p}}
\eeq
for $\text{B}^{8-p}$ the base of a Ricci flat cone (for instance $\text{B}^{8-p}=\text{S}^{8-p}$). The case of $p=8$ is different as while the solution is singular at such a loci, the metric neither blows up nor tends to zero so a D8 brane does not bound the solution ($p=7$ will not be relevant to our analysis). The Op plane singularity  (for $p\neq 7$) on the other hand yields
\beq
ds^2_{\perp}\propto dr^2+ \alpha_0^2 ds^2(\text{B}^{8-p}),~~~ h_p\propto r,
\eeq
for $\alpha_0$ some constant. Our task then is to first establish which of these behaviours can be realised by the class of solutions in this section, then whether any of these behaviours can coexist in the same local solution. Let us reiterate: D-brane and O-plane singularities do not exhaust the possible physical singularities, indeed we will find a more complicated object later in this section which we will describe when it becomes relevant. \\
~~\\
The most obvious thing one can try is to fix $F_0=h'''=0$, it is not hard to see that one then has $H_3=0$ and $F_4$ becomes purely electric. One can integrate $h$ as
\beq
h=  c_1+ r c_2+\frac{1}{2} \tilde{k}r^2
\eeq
and then upon making the redefinitions
\beq
c_1=\frac{\tilde{k} L^4+c_2^2 \pi^2}{2 \tilde{k} \pi^2},~~~~~ -r=\frac{c_2}{\tilde{k}^2}+\frac{L^2}{\pi}\sinh x
\eeq
one finds the solution is mapped to 
\begin{align}
\frac{ds^2}{L^2}&= \bigg(\cosh^2x ds^2(\text{AdS}_3)+ dx^2\bigg)+ 4 ds^2(\mathbb{CP}^3),~~~~ e^{-\Phi}=\frac{\tilde{k}}{2 L},\nn\\[2mm]
F_2&= 2 \tilde{k}J_2,~~~~~ F_4=\frac{3}{2} \tilde{k} L^2 \cosh^3 x \text{vol}(\text{AdS}_3)\wedge dx.
\end{align}
This is of course AdS$_4\times \mathbb{CP}^3$, dual to ${\cal N}=6$, U(N)$_{\tilde{k}}\times$U(N)$_{-\tilde{k}}$  Chern-Simons matter theory (where $N= 2 \tilde{k} L^4$) \cite{Aharony:2008ug}. Thus there is only one local solution when $F_0=0$ and it is an AdS$_4$ vacua preserving twice the supersymmetries of generic solutions within this class. This is  the only regular solution preserving (6,0) supersymmetry.\\
~~\\
Next we consider the sort of physical singularities the metric and dilaton in \eqref{eq:neq6nssector} can support for $F_0\neq 0$. At the loci of such singularities the space terminates so the interval spanned by $r$ becomes bounded at one end. We shall use diffeomorphism invariance to assume this bound is at $r=0$. 

First off $\Delta_1=2 h h'' -( h')^2$ appears in the metric and dilaton where one would expect the warp factor of a co-dimension 7 source to appear. Thus if $\Delta_1$ has an order 1 zero at a loci where $h,h''$ have no zero one has the behaviour of O2 planes extended in AdS$_3$ at the tip of a G$_2$ cone over $\mathbb{CP}^3$. We now choose the loci of this O2 plane to be $r=0$, meaning that the constant part of $\Delta_1$ has to vanish which forces
\beq
\text{O2 at}~ r=0:~ h=c_1+c_2 r + \frac{c_2^2}{4 c_1}r^2+\frac{1}{3!}c_3 r^3,~~~c_1,c_2,c_3 \neq 0,\label{eq:O2}
\eeq
where  $r\in \mathbb{R}^{\pm}$ when  $\text{sign}(c_1 c_4)=\pm 1$.

Another type of singularity this solution is consistent with is a D8/O8 system of world volume AdS$_3\times \mathbb{CP}^3$. Such a singularity is characterised by O8 brane like behaviour in the metric and dilaton. We realise this behaviour by choosing $h$ such that $(h'',\Delta_1)$ both have an order 1 zero at a loci where $h$ has no zero. After using diffeomorphism invariance to place the D8/O8 at $r=0$ this is equivalent to taking  $h$ as
\beq
\text{D8/O8 at}~ r=0:~ h=c_1+ \frac{1}{3!}c_2 r^3,~~~ c_{1,2}\neq 0.,\label{eq:D8O8}
\eeq
where $r\in \mathbb{R}^{\pm}$ for $\text{sign}(c_1 c_2)=\pm 1$.

We are yet to find a D brane configuration, given that we have a $\mathbb{CP}^3$ factor the obvious thing to naively aim for is D2 branes at the tip of a G$_2$ cone  similar to the O2 plane realised above. However the warp factor of a D2 brane blows up like $\lim_{r\to 0}r^{-5}$ its loci, and this is not possible to achieve for \eqref{eq:neq6nssector} such that $h$ is an order 3 polynomial. It is however possible to realise a more exotic object: It is well known that if one takes $d=11$ supergravity on the orbifold $\mathbb{R}^{1,6}\times\mathbb{C}^2/\mathbb{Z}_{\tilde{k}}$  then reduces on the Hopf fibre of the Lens space (equivalently squashed 3-sphere) inside $\mathbb{C}^2/\mathbb{Z}_{\tilde{k}}$ one generates a D6 brane singularity in type IIA. One can generate the entire flat space D6 brane geometry by replacing $\mathbb{C}^2/\mathbb{Z}_{\tilde{k}}$ in the above with a Taub-Nut space and likewise reducing to IIA. One can perform an analogous procedure for $\mathbb{R}^{1,2}\times\mathbb{C}^4/\mathbb{Z}_{\tilde{k}}$, reducing this time on the Hopf fibration (over $\mathbb{CP}^3$) of a squashed 7-sphere. The resulting solution in IIA takes the from
\beq
ds^2= \frac{\sqrt{r}}{{\tilde{k}}} ds^2(\text{Mink}_3)+ \frac{1}{4 {\tilde{k}} \sqrt{r}}\bigg(dr^2+ 4 r^2 ds^2(\mathbb{CP}^3)\bigg),~~~e^{-\Phi}=\frac{{\tilde{k}}^{\frac{3}{2}}}{r^{\frac{3}{4}}},~~~F_2= 2{\tilde{k}} J_2,~~~r\geq 0 \label{eq:novelsingularity},
\eeq
and is singular at $r=0$. Notice that the $r$ dependence of the dilaton and metric is the same as one gets at the loci of flat space D6 branes, but the co-dimensions no longer span a regular cone as they do in that case, indeed $dr^2+ c r^2 ds^2(\mathbb{CP}^3)$ is only Ricci flat for unit radius round $\mathbb{CP}^3$ when $c=\frac{2}{5}$. It is argued in \cite{Aharony:2008ug} that the singularity in \eqref{eq:novelsingularity} corresponds to a coincident combination of a KK monopole and ${\tilde{k}}$ D6 branes (T-dual of an $(1,{\tilde{k}})-5$ brane) that partially intersect another KK monopole. For simplicity we shall refer to this rather complicated composite object as a $\widetilde{\text{D6}}$ brane.  We can find the behaviour of this object, now extended in AdS$_3$, within this class -- assuming it is located at $r=0$ one need only tune
\beq
\widetilde{\text{D6}}~\text{at}~r=0:~ h= r^2(c_1+c_2 r),~~~~c_{1,2}\neq 0,\label{eq:tildeD6}
\eeq
with the caveat that as this only exists when $F_0\neq 0$, we can no longer lift to $d=11$. Again $r\in \mathbb{R}^{\pm}$ for $\text{sign}(c_1 c_2)=\pm 1$.

The above exhausts the \textbf{physical} singularities we are able to identify,  we do however find one final further singularity. By tuning $h= c r^3$ for $r,c>0$, the metric and dilaton then become
\beq
ds^2 =\frac{\pi}{2\sqrt{3}r}\bigg(4 r^2\left(ds^2(\text{AdS}_3)+ ds^2(\mathbb{CP}^3)\right)+3 dr^2\bigg),~~~e^{-\Phi}= \frac{3 \sqrt{2}3^{\frac{1}{4}}c}{\sqrt{\pi}}\sqrt{r}\label{eq:weirdneq6singularity}
\eeq
which is singular about $r=0$ in a way we do not recognise.\\
~~\\
All the previously discussed physical singularities  bound the interval spanned by $r$ at one end. In order to have a true AdS$_3$ vacuum we need to bound a solution between 2 of them separated by a finite proper distance. However assuming that the space starts at $r=0$ with any of an O2, D8/O8 or $\widetilde{\text{D6}}$ singularity, we find that that none of the warp factors appearing in metric or dilaton \eqref{eq:neq6nssector}, (ie  $(h,\Delta_1,h'')$) either blow up or vanish until $r\to \infty$. For each case (assuming $r\geq 0$ for simplicity) the metric and dilaton as $r\to \infty$ tend to \eqref{eq:weirdneq6singularity} . By computing the curvature invariants it is possible to show that the metric is actually flat at this loci, hence it tends to $\text{Mink}_{10}$, however as $e^{\Phi} \to \infty$ the solution is still singular. Worse still the singularity as $r\to \infty$ is at infinity proper distance from $r=0$, so in all cases the internal space is semi infinite.

Naively one might now conclude that there are no $\mathfrak{osp}(6|2)$ AdS$_3$ vacua with bounded internal space (true vacua), however as we will show explicitly in section \ref{sec:global}, this is not the case. The missing ingredient is the inclusion of D8 branes on the interior of $r$ which allow one to glue local solutions of $h''''=0$, depending on different integration constants, together.

\subsection{ Analysis of $\mathfrak{osp}(5|2)$ vacua}   \label{sec:osp52}
The $\mathfrak{osp}(5|2)$ AdS$_3$ solutions are given by the class of section \ref{ses:classification} for $u' \neq 0$, one can use diffemorphism invariance to fix $u=r$ for such solutions without loss of generality. The resulting NS sector takes the form
\begin{align}
\frac{ds^2}{2\pi}&= \frac{|hr|}{\sqrt{\Delta_1}}ds^2(\text{AdS}_3)+\frac{\sqrt{\Delta_1}}{4|r|}\bigg[ \frac{2}{ |h''|} \bigg(ds^2(\text{S}^4)+ \frac{1}{\Delta_2 }(Dy_i)^2\bigg)+\frac{1}{|h| }dr^2\bigg],\nn\\[2mm]
e^{-\Phi}&=\frac{ |h''|^{\frac{3}{2}}\sqrt{|r|}\sqrt{\Delta_2 }}{2\sqrt{\pi}\Delta_1^{\frac{1}{4}}},~~~~\Delta_1=2 h h'' r^2-(r h'- h )^2,~~~~\Delta_2=1+\frac{2 h' }{r h''},~~~~H=dB_2,\nn\\[2mm]
 B_2&= 4\pi\bigg[\left(-(r-k)+\frac{r h'-h }{r h''}\right) J_2+ \frac{1}{2h''}\left(\frac{h}{r}+\frac{h h''-2 (h')^2}{2h'+r h''}\right)\left(J_2- \tilde J_2\right)\bigg],
\end{align}
while the $d=10$ RR fluxes are then  given by
\begin{align}
F_0&=-\frac{1}{2\pi}h''',~~~~~F_2= B_2 F_0+  2(h''-(r-k) h''')J_2,\nn\\[2mm]
F_4&=\pi d\left(h'+\frac{hh''r(rh'+ h )}{\Delta_1}\right)\wedge \text{vol}(\text{AdS}_3)+B_2 F_2-\frac{1}{2}B_2\wedge B_2 F_0\nn\\[2mm]
& -4\pi\bigg[(2h'+(r-k)(-2 h''+(r-k) h'''))J_2\wedge J_2+ d\left(\frac{h }{r}\text{Im}\Omega_3\right)\bigg],
\end{align}
where  $h$ is defined as before and $k$ is a constant.

\subsubsection{Local solutions and regularity}
In this section we will explore the physically distinct local $\mathfrak{osp}(5|2)$ solution that follow from solving  $h'''=-2\pi F_0$ in  various ways.\\
~~\\
Let us begin by commenting on the massless limit $F_0=h'''=0$: Unlike the class of $\mathfrak{osp}(6|2)$ solutions the result is no longer locally AdS$_4\times \mathbb{CP}^3$ - it is instructive to lift the class to $d=11$.  We find the metric of the solution can be written as\footnote{Note that to get to this form one must rescale the canonical 11'th direction, ie if this is $z$ and $L^i_{1,2}$ are defined as in \eqref{eq:leftinvariantforms} then $\phi_2=\frac{2}{h''}z$}
\begin{align}
\label{eq:elevenmetric}
	\frac{ds_{11}^2}{2^{\frac{1}{3}}\pi^{\frac{2}{3}}|h''|}&=\Delta_2^{\frac{1}{3}}\bigg[\frac{|h||r|^{\frac{4}{3}}}{\Delta_1^{\frac{2}{3}}}ds^2(\text{AdS}_3)+\frac{\Delta_1^{\frac{1}{3}}}{ |r|^{\frac{2}{3}}}\bigg(\frac{1}{4|h|}dr^2+\frac{2}{|h''|}ds^2(\widehat{\text{S}}^7)\bigg)\bigg],\nonumber\\[2mm]
	ds^2(\widehat{\text{S}}^7)&=\frac{1}{4}\bigg[ds^2(\text{S}^4)+\frac{1}{\Delta_2}\left(L_2^i- \cos^2 \left(\frac{\alpha}{2}\right) L_1^i\right)^2\bigg],
\end{align}
where $L_{1,2}^i$  are two sets of SU(2) left invariant forms defined as in appendix \eqref{eq:leftinvariantforms}, so that the internal space is a foliation of an SP(2)$\times$SP(1) preserving squashed 7-sphere over an interval.  This enhancement of symmetry is also respected by the $d=11$ flux, to see this we need to define the SP(2)$\times$SP(1) invariant form on this squashed 7-sphere, fortuitously these were already computed in \cite{Legramandi:2020txf}, they are
\beq
\Lambda_3^0=\frac{1}{8}(L_2^1+\mathcal{A}^1)\wedge(L_2^2+\mathcal{A}^2)\wedge(L_2^3+\mathcal{A}^3),~~\tilde{\Lambda}_3^0=\frac{1}{8}(L_2^i+\mathcal{A}^i)\wedge(d\mathcal{A}^i+\frac{1}{2}\epsilon^i_{jk}\mathcal{A}^j\wedge\mathcal{A}^k),
\eeq
where $\mathcal{A}^i=- \cos^2 \left(\frac{\alpha}{2}\right) L_1^i$, and their exterior derivatives. One can then show that the $d=11$ flux decompose as
\beq
\frac{G_4}{\pi}=d\left(h'+\frac{rhh''(rh'+h)}{\Delta_1}\right)\wedge \text{vol}(\text{AdS}_3)+4d\bigg(\frac{2(r(h')^2-h(h'+rh''))}{r(2h'+rh'')}\Lambda_3^0+\frac{ h}{r}(\Lambda_3^0-\tilde{\Lambda}_3^0)\bigg).
\eeq
For such solutions we can in general integrate $h'''=0$ in terms of an order 2 polynomial. As we shall see shortly it is possible to bound $r$ at one end of the space in several physical ways, but when $F_0=0$ it always remains semi-infinite. Given that the massless limit of ${\cal N}=(6,0)$ is always locally AdS$_4\times \mathbb{CP}^3$ in IIA, it is reasonable to ask whether the massless solutions here approach this asymptotically. Such a solution preserving  ${\cal N}=(8,0)$ was found on this type of squashing of the 7-sphere  in \cite{Legramandi:2020txf} and can be interpreted as a holographic dual to a surface defect. In this case, as $r\to \infty$  the the curvature invariants \eqref{eq:elevenmetric} all vanish, ( for instance $R\sim r^{-\frac{2}{3}}$). This makes the behaviour at infinite $r$ that of Mink$_{11}$, so such an interpretation is not possible here.\\
~~\\
Let us now move back to IIA and focus on more generic solutions: By studying the zeros of $(r ,~h,~h'',~\Delta_1,~\Delta_2)$ we are able to identify  a plethora of  boundary behaviours for $\mathfrak{osp}(5|2)$ solutions. The vast majority we are able to identify as physical and most exist for arbitrary values of $F_0$. We already used up translational invariance of this class to align $u=r$ so we can non longer assume that $r=0$ is a boundary of solutions in this class, rather we must consider possible boundaries at  $r=0$ and $r= r_0$ for $r_0\neq 0$ separately.

We have two physical boundary behaviours that only exist for $F_0 \neq 0$: The first of these is a regular zero for which the warp factors of AdS$_3$ and S$^4$ become constant while the $(r,\text{S}^2)$ directions approach the origin of  $\mathbb{R}^3$ in polar coordinates. This is given by tuning
\beq
\text{Regular zero at} ~ r=0:~~~~h=c_1 r+\frac{1}{3!}c_2r^3,~~~c_{1,2}\neq 0,~~~ \text{sign}(c_1 c_2)=1
\eeq
where one can take either of $r\in \mathbb{R}^{\pm}$ . This is the only regular boundary behaviour that is possible.

Next it is possible to realise a fully localised O6 plane of world volume  $(\text{AdS}_3,~ \text{S}^4)$ at $r=r_0$ by tuning
\beq
\text{O6 plane at} ~ r=r_0:~~~~h=c_1 r_0+c_1 (r- r_0)+\frac{1}{3!}c_2(r-r_0)^3,~~~ c_1,c_{2},r_0 \neq 0\label{eq:neq5O6}.
\eeq 
The domain of $r$ in this case depends more intimately on the tuning of  $c_1,c_2,r_0$ than we have thus far seen: When $r_0<0$   one has $r\in (-\infty,~r_0]$ for $\text{sign}(c_1 c_2)=1$ while for $\text{sign}(c_1 c_2)=-1$ one finds that $ r_0\leq r \leq r_1<0$ for $r_1=r_1( c_1,c_2)$. Conversely for $r_0>0$, $\text{sign}(c_1 c_2)=1$ implies $r\in [r_0,~\infty)$ while $\text{sign}(c_1 c_2)=-1$ implies $ 0<r_1\leq r \leq r_0$.


The remaining boundary behaviour exist whether $F_0$ is non trivial or not: We find the behaviour of D6 branes extended in  $(\text{AdS}_3,~ \text{S}^4)$  by tuning
\beq
\text{D6 brane at}~r=0:~~~~h=c_1 r+ \frac{1}{2}c_2 r^2+\frac{1}{3!}c_3 r^3,~~~~c_{1,2}\neq 0.\label{eq:neq5D6}
\eeq
When $\text{sign}(c_1c_2)=\pm 1$ $r=0$ is a lower/upper bound. Given this, $r$ is also bounded from above/below when $\text{sign}(c_1 c_3)=- 1$  and is semi-infinite for $\text{sign}(c_1 c_3)=1 $ and $c_3=0$.

As with the $\mathfrak{osp}(6|2)$ class it is possible to realise a $\widetilde {\text{D}6}$ singularity (see the discussion below \eqref{eq:D8O8}), this time at $r=r_0$ by tuning 
\beq
\widetilde {\text{D}6}~\text{brane at}~r=r_0:~~~~h= \frac{1}{2}c_1 (r-r_0)^2+\frac{1}{3!}c_2 (r-r_0)^3,~~~~r_0,c_2\neq 0,~~~~c_2\neq -3 \frac{c_1}{r_0}.\label{eq:neq5tildeD6}
\eeq
For $\text{sign}(r_0 c_1 c_2)=1$ the domain of $r$ is semi infinite bounded from above/below when  $\text{sign}(r_0 )=\mp 1$. When $\text{sign}(r_0 c_1 c_2)=-1$ we find that $r$ is bounded between $r_0$ and some constant $r_1=r_1(r_0,c_1,c_2)$. Given the later behaviour, when $\text{sign}(r_0)=\pm 1$ one finds that $r$ is strictly positive/negative with $r_0$  the upper/lower of the 2 bounds when $|c_2 r_0|>3 |c_1|$ and the lower/upper when $|c_2 r_0|<3 |c_1|$. 

Next we find the behaviour of an O4 plane extended in AdS$_3\times$S$^2$ by tuning
\beq
\text{O4}~\text{plane at}~ r=r_0:~~~~ h=\frac{1}{2}c_1\big( r_0^2- r_0(r-r_0)+  (r-r_0)^2\big)+\frac{1}{3!}c_2(r-r_0)^3,~~~c_1,r_0\neq 0,~~~ c_2 \neq-\frac{3c_1}{r_0}.\label{eq:neq5O4}
\eeq
In this solution the domain of $r$ has the same qualitative dependence on the signs of $c_1,c_2,r_0$ and whether $|c_2 r_0|>3 |c_1|$ or $|c_2 r_0|<3 |c_1|$ as the previous example, though  the precise value of $r_1(c_1,c_2,r_0)$ is different.

Likewise we find the behaviour of an O2 plane extended in AdS$_3$ and back-reacted on a G$_2$ cone whose base is round $\mathbb{CP}^3$, this is achieved by tuning
\beq
\text{O2}~\text{plane at}~ r=r_0:~~~~~h= 2r_0c_1+\frac{1}{2 }c_1(r-r_0)^2+ \frac{1}{3!}c_2 (r-r_0)^3,~~~~r_0,c_1\neq 0,~~~ c_2 \neq-\frac{3c_1}{r_0},\label{eq:neq5O2}
\eeq
where the domain of $r$ is qualitatively related to the parameters as it was for the $\widetilde {\text{D}6}$.

Finally we find the behaviour of an O2' plane extended in AdS$_3$ and back-reacted on a G$_2$ cone whose base is a squashed $\mathbb{CP}^3$  (ie $4ds^2(\text{B}^6)=2ds^2(\text{S}^4)+ds^2(\text{S}^2)$) at $r=r_0$ by tuning
\beq
\text{O2'}~\text{plane at}~ r=r_0:~~~~~h= 2c_1r_0^2+2 c_1 b(r-r_0)+\frac{1}{2 }c_1(b-1)^2(r-r_0)^2+ \frac{1}{3!}c_2 (r-r_0)^3,\label{eq:neq5O2d}
\eeq
where we must additionally impose $r_0,c_1\neq 0$. This gives behaviour similar to the O2 plane, ie $e^{2A}\sim (r-r_0)^{-\frac{1}{2}}$ with the rest of the warp factors scaling as the reciprocal of this. However in general  $e^{2(C-D)}=\left(\frac{b+1}{b-1}\right)^2$ at leading order about $r=r_0$ and the internal space only spans a Ricci flat cone for $e^{2(C-D)}=1,2$, with the former yielding \eqref{eq:neq5O2}. As such for the O2' plane we must additionally tune
\beq
\left(\frac{b+1}{b-1}\right)^2=2,\label{eq:O2'cond}
\eeq
which has two solutions $b_{\pm}=3\pm 2\sqrt{2}$ and we must have $c_2 r_0\neq- 12 b_{\pm}$. Again the domain of $r$ has the same qualitative dependence as the $\widetilde{\text{D}}6$, though this time the relevant equalities that determine whether $r_0$ is an upper or lower bound are $|c_2 r_0|>12 b_{\pm} |c_1|$ or $|c_2 r_0|<12 b_{\pm} |c_1|$. This exhausts the the physical singularities we have been able to identify.\\
~~\\
As with the case of $\mathfrak{osp}(6|2)$ vacua we have been able to identify several local solution for which the domain of $r$ is semi infinite. For these  the metric as $r\to\pm \infty$ is once again flat, but at infinite distance and with a non constant dilaton. For the $\mathfrak{osp}(5|2)$ solutions however it is possible to bound the majority of the solutions for suitable tunings of the parameters on which they depend - this necessitates $F_0\neq 0$. A reasonable question to ask then is which physical singularities can reside in the same local solution? There are actually 7 distinct local solutions bounded between two physical singularities, we provide details of these in table \ref{table:1}. Note that the solution with regular zero is unbounded while the D6 solution can only be bounded by a singularity of the type given in \eqref{eq:neq5O2d}, but without \eqref{eq:O2'cond} being  satisfied so is thus non-physical.\\
~~\\
In this section, and the preceding one  we have analysed the possible local solutions preserving ${\cal N}=(5,0)$ and $(6,0)$ supersymmetry that follow from various tunings of the (local) order 3 polynomial $h$. We found many different possibilities, many of which can give rise to a bounded interval in the $(5,0)$ case, but non of which do in the (6,0) case. This is not the end of the story, in this section we have assumed that $F_0$ is a constant which excludes the presence of D8 branes along the interior of the interval. In the next section we shall relax this assumption allowing for much wider classes of global solution, and in particular $(6,0)$ solutions with bounded internal space.
\begin{center}
\begingroup
\renewcommand{\arraystretch}{1.7}
\begin{sidewaystable}
\centering
\begin{tabular}{||c | c | c  ||} 
 \hline
 Bound at $r=r_0$  &  Bound at $r=\tilde{r}_0$  & Additional tuning and comments\\ [0.5ex] 
 \hline\hline
O6~\eqref{eq:neq5O6}&O4 $\lvert$ O2 $\lvert$ O2'&  $\begin{array}{c}c_1=\frac{(\tilde{r}_0-r_0)^3(r_0^2+4 r_0 \tilde{r}_0-8 \tilde{r}_0^2)}{72 \tilde{r}_0^3}c_2,\\ r_0=\frac{(\sqrt{3}\sqrt{11+8 \sqrt{b_0}}-1)\tilde{r}_0}{2}\end{array}
$~~~~ $b_0=0~\lvert~ 1~\lvert~ 2$ \\
\hline
D6~\eqref{eq:neq5D6}& \text{Non-physical}& $\begin{array}{c}\tilde{r}_0:~\text{Middle of 3 real roots of }\\8 c_3^2 \tilde{r}_0^3+36 c_2 c_2 \tilde{r}_0^2+(27c_2^2+72 c_1 c_3)\tilde{r}_0+72 c_1 c_2=0\end{array}$,~~~\eqref{eq:neq5O2d} like  but for  $e^{2C-2D}>2$\\
\hline
$\widetilde {\text{D}6}$~\eqref{eq:neq5tildeD6}&O2'&$\begin{array}{c}c_1=\frac{5-4\sqrt{2}-\sqrt{120+86 \sqrt{2}}}{21}c_2 r_0\\ \tilde{r}_0=\frac{\sqrt{4+3 \sqrt{2}}r_0}{2}\end{array}$~\\
\hline
O4~\eqref{eq:neq5O4}&O2 &$\begin{array}{c}c_1=  c_2 b_- r_0,\\ \tilde{r}_0= b_+ r_0\end{array}$,~~~~$\begin{array}{c}b_{\pm}:~\pm\text{tive real roots of}\\13 b_+^3-29 b_+^2+10b_+=6\\1287 b_-^3+610 b_-^2-1287  b_-+144=0\end{array}$ \\
\hline
O4~\eqref{eq:neq5O4}&O2' &$\begin{array}{c}c_1=  c_2 b_- r_0,\\ \tilde{r}_0= b_+ r_0\end{array}$,~~~~$\begin{array}{c}b_{\pm}:~\pm\text{tive real roots of}\\2 b_+^6-8 b_+^5+40 b_+^3-49 b_+^2+6 b_+=9\\81 b_-^6-18b_-^5+37 b_-^4+12 b_-^3-93 b_-^2+54 b_-=9\end{array}$ \\
\hline
O2~\eqref{eq:neq5O2}&O2'&$\begin{array}{c}c_1=  c_2 b_- r_0,\\ \tilde{r}_0= b_+ r_0\end{array}$,~~~~$\begin{array}{c}b_{\pm}:~\pm\text{tive real roots of}\\8 b_+^6-16b_+^5-8 b_+^4+224 b_+^3-209 b_+^2+26 b_+=169\\711 b_-^6-2388 b_-^5+2858b_-^4-972 b_-^3-705b_-^2+648 b_-=144\end{array}$\\
 \hline\hline
\end{tabular}
\caption{A list of distinct local $\mathfrak{osp}(5|2)$ solutions with bounded internal space. Note that we do not include solutions with the singularities at $r=r_0$ and $r=\tilde{r}_0$ inverted, as they are physically equivalent.}
\label{table:1}
\end{sidewaystable}
\endgroup

\end{center}

\section{Global solutions with interior D8 branes}\label{sec:global}
In this section we show that it is possible to glue the local solutions of section \ref{sec:local} together with D8 branes placed along the interior of the interval spanned by $r$. This opens the way for constructing many more global solutions with  bounded internal spaces.\\
~\\
In the previous section we studied what  types of local ${\cal N}=(5,0)$ and $(6,0)$ it is possible to realise with various tunings of the order 3 polynomial $h$. The assumption we made before was that $F_0=$ constant was fixed globally, but this is not necessary, all that is actually required is that $F_0$ is piecewise constant. A change in $F_0$ gives rise to a delta function source as
\beq
dF_0= \Delta F_0 \delta(r-r_0) dr,
\eeq 
where $\Delta F_0$ is the difference between the values $F_0$ for $r>r_0$ and $r<r_0$.  D8 branes give rise to a comparatively mild singularity for which the Bosonic fields neither blow up nor tend to zero so do not represent a boundary of a solution, indeed the solution continues passed them unless they appear coincident to an O8 plane. As one crosses a D8 brane the metric and dilaton and NS 3-form are continuous, but the RR sector can experience a shift.  To accommodate such an object within the class of solutions of section \ref{ses:classification}  one needs to do so in terms of $h$, so it can lie along $r$. If we wish to place a D8 at $r=r_0$  one should have
\beq
h''''=- \Delta N_8 \delta(r-r_0),~~~~~ \Delta N_8= 2\pi \Delta F_0,
\eeq
where $\Delta N_8$ is the difference in D8 brane charge between  $r>r_0$ and $r<r_0$ and is the charge of the D8 brane source at $r=r_0$.
While the conditions that the NS sector should be continuous amounts to demanding the continuity of
\begin{align}\label{eq:continuity}
{\cal N}=(6,0)&: ~~~~(h,(h')^2, h''),\\[2mm]
{\cal N}=(5,0)&: ~~~~(h,h', h'')\nn,
\end{align}
recall that $u''=0$ is a requirement for supersymmetry so $u$ cannot change as one crosses the D8. The source corrected Bianchi identities in general take the form
\beq
(d-H\wedge) f_+=  \frac{1}{2\pi}\Delta N_8\delta(r-r_0) dr\wedge e^{{\cal F}}~~~~\Rightarrow~~~~~ d\hat f_+= \frac{1}{2\pi}\Delta N_8\delta(r-r_0) dr\wedge e^{2\pi \tilde{f}}
\eeq
where ${\cal F}=  B_2+ 2\pi \tilde{f}$ for $\tilde{f}$ a world volume gauge field on the D8 brane and where $\hat f_+$ are the magnetic Page fluxes, \eqref{eq:pagefluxess} for the solution at hand. If $\tilde{f}$ is non zero  then the D8 is actually part of a bound state involving every brane whose flux receives a source correction in $d\hat f_+$. 
Recalling the Bianchi identities \eqref{eq:bianchis}, we have for the specific case at hand that
\beq
 d \hat f_n=\frac{1}{2\pi}(4\pi )^n\frac{1}{n!}(r-k)^n \Delta N_8 \delta(r-r_0) dr\wedge J_2^n
\eeq
We thus see that it is consistent to set the world volume flux on the D8 brane to zero by tuning $r_0$, ie 
\beq
\tilde{f}=0~~~\Rightarrow~~~ r_0=k. 
\eeq
This means only $F_0$ receives a delta function source, the rest vanishing as   $(r-r_0)^n \delta(r-r_0)\to 0$ for $n>0$. So we have shown that it is possible to place D8 branes, that do not come as a bound state, at $r=k$ and solve the Bianchi identities - but how should one interpret this? The origin of $k$ in our classification was as an integration constant, but one can view it as the result of performing a large gauge transformation, ie shifting the NS 2-form  by a form that is closed but not  exact  as
$B_2 \to  B_2+ \Delta B_2$ such that $b_0=\frac{1}{(2\pi)^2}\int_{\Sigma_2} B_2$ is quantised over some 2-cycle $\Sigma_2$. Clearly squashed $\mathbb{CP}^3$ contains an S$^2$ which $B_2$ has support on, that of the fiber, one finds
\beq
\frac{1}{(2\pi)^2}\int_{\text{S}^2} B_2-\frac{1}{(2\pi)^2}\int_{\text{S}^2} B_2\bigg\lvert_{k=0} = \frac{k}{4\pi}\int_{\text{S}^2}\text{vol}(\text{S}^2)= k, 
\eeq
so provided $k$ is quantised\footnote{Note: usually one means integer by quantised, by in the presence of fractional branes, such as in ABJ \cite{Aharony:2008gk} which shares a $\mathbb{CP}^3$ in its internal space, it is possible for parameters such as $k$ to merely be rational.} its addition to the minimal potential giving rise to the NS 3-form is indeed the action of a large gauge transformation. The key point is that because $k$ follows from a large gauge transformation, it does not need to be fix globally, indeed in many situations where $B_2$ depends on a function of the internal space it is necessary to perform such large gauge transformations as one move through the internal space to bound $b_0$ to with some quantised range.  We conclude that the Bianchi identities are consistent with placing D8 branes at quantised $r=k$  loci provided that they are accompanied by the appropriate number of large gauge transformations of the NS 2-form.\\
~~\\
Of course to be able claim a supersymmetric vacua the sources themselves need to have a supersymmetric embedding, further if this is the case the integrability arguments of \cite{Martucci:2011dn,Prins:2013koa}  imply that the remaining type II equations of motion of the Bosonic supergravity are implied by the Bianchi identities we have already established hold. The existence of a supersymmetric brane embedding can  be phrased in the language of (generalised) calibrations \cite{Gutowski:1999tu}: A source extended in  AdS$_3$ and wrapping some  n-cycle $\Sigma$ is supersymmetric if it obeys the following condition 
\beq
\Psi^{\text{cal}}_n=8\Psi_+\wedge e^{-{\cal F}}\bigg\lvert_{\Sigma}=\sqrt{\det(g+{\cal F})} d\Sigma \bigg\lvert_{\Sigma}\label{eq:calibration}
\eeq
where $\Psi_+$ is the bi-linear appearing in  \eqref{eq:so5bilinears}, $g$ is the metric on the internal space and where the pull back onto ${\Sigma}$ is understood.  For a D8 brane placed along $r$ we take $\tilde{f}=0$,  $d\Sigma=\sin^3\alpha d\alpha\wedge \text{vol}(\text{S}^3)\wedge \text{vol}(\text{S}^2)$ and $B_2$ defined as in \eqref{eq:summerystart} --  we find 
\begin{align}
\Psi^{\text{cal}}_6=\frac{\pi^3}{2\sqrt{2}u^3h''\sqrt{h h''}\sqrt{\Delta_2}}\bigg[&u\left(h(2u-r_ku')-r_ku h'\right)\left(2h\left(u+r_ku'\right)- u\left(2h'-h''r_k\right)r_k\right)\nn\\[2mm]
&+2\cos^2\alpha u' \Delta_1 r_k^3\bigg]  d\Sigma,\nn\\[2mm] 
\sqrt{\det(g+{\cal F})} d\Sigma \bigg\lvert_{\Sigma}=\frac{\pi^3}{8(u h'')^{\frac{3}{2}}\sqrt{\Delta_2 }}&\sqrt{2(h-h' r_k)(u-u' r_k)+ u h'' r_k^2}\nn\\[2mm]&\times \left(2h(u+u' r_k)- u r_k(2h'-h'' r_k)\right)d\Sigma
\end{align}
where we use the shorthand $r_k=r-k$. It is simple to then show that \eqref{eq:calibration} is indeed satisfied for a D8 brane at  $r=k$, and so supersymmetry is preserved.\\
~\\
In summary we have shown that D8 branes can be placed along the interior of $r$ at the loci $r=k$, for quantised $k$, without breaking supersymmetry provided an appropriate large gauge transformation of the NS 2-form is performed.  This allows one to place a potentially arbitrary number of D8 branes along $r$ and use them to glue the various local solutions of section \ref{sec:local} together provided that the continuity of \eqref{eq:continuity} holds across each D8. In the next section we will explicitly show this in action with 2 examples.

\subsection{Some simple examples with internal D8 branes}  
\begin{figure}
\centering
\includegraphics[scale=0.35]{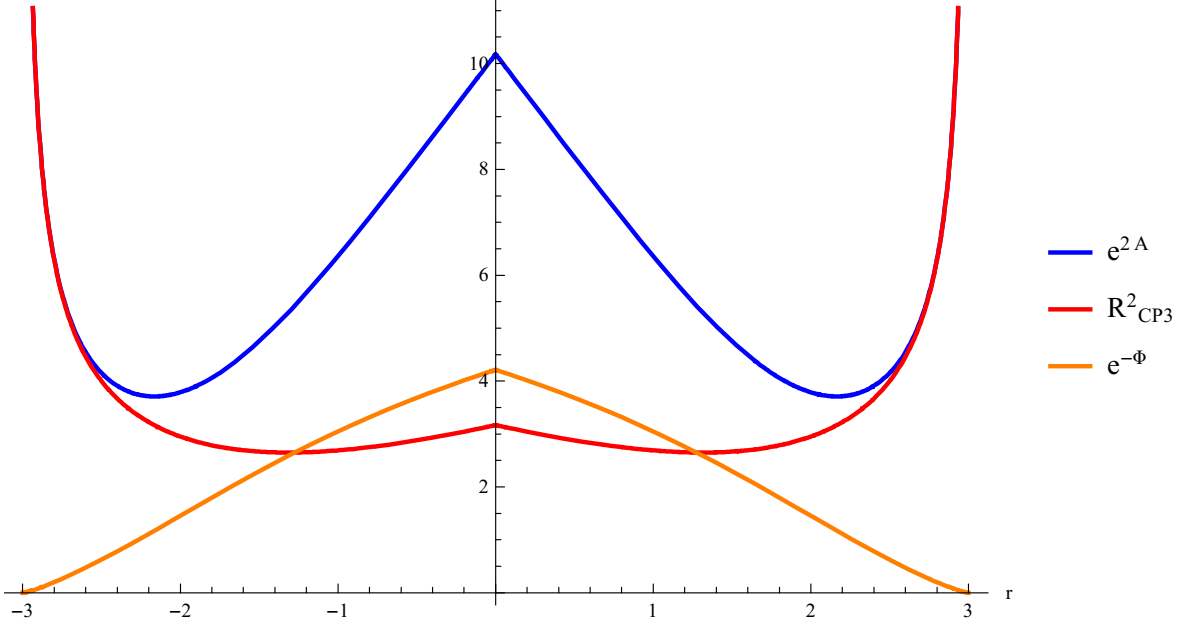}
\caption{Plot of the warp factors in the metric and dilaton for an ${\cal N}=(6,0)$ solution with D8 at $r=0$ bounded between D8/O8 singularities at $r=\pm 3 $ with the remaining constants in $h$ tuned as $c_1=2$, $c_2=5$}
\label{fig1}
\end{figure}

In this section we will construct two global solutions with interior D8 branes and bounded internal spaces, one preserving each of ${\cal N}=(6,0)$ and $(5,0)$ supersymmetry. Let us stress that this only scratches the surface of what is possible, we  save a more thorough investigation for forthcoming work \cite{USandYolanda}.\\
~\\
We shall first construct a solution preserving ${\cal N}=(6,0)$, meaning that we need to impose the continuity of $(h,(h')^2, h'')$ as we cross a D8. Probably the simplest thing one can do  is to place a stack of D8 branes at the origin $r=0$  and bound $r$ between  D8/O8 brane singularities which are symmetric about this point. As such we can take $h$ to be globally defined as 
\beq
h=\left\{\begin{array}{l}-c_1- \frac{c_2}{3!}(r+r_0)^3~~~~ r<0\\[2mm]-c_1- \frac{c_2}{3!}(r_0-r)^3~~~~ r>0\end{array}\right.
\eeq
This bounds the interval to   $-r_0<r< r_0$ between D8/O8 singularities at $r=\pm r_0$ and gives rise to a source for the $F_0$ of charge  $2c_2$, ie
\beq
h''''=  -2 c_2 \delta(r)~~~~\Rightarrow~~~~dF_0=  2 c_2\frac{1}{2\pi} \delta(r) dr
\eeq
The form that the warp factors and metric take for this solution is depicted in figure \ref{fig1}.
Given the Page fluxes in \eqref{eq:pagefluxess} (with $u=1$), and that we simply have round $\mathbb{CP}^3$ for this solution for which we can take  $\int_{\mathbb{CP}^n}J_2^n= \pi^n$, it is a simple matter to compute the Page charges of the fluxes over the $\mathbb{CP}^n$ sub-manifolds of $\mathbb{CP}^3$. By tuning 
\beq
c_1= N_2- \frac{N_5^3N_8}{6},~~~~c_2= N_8,~~~~r_0=N_5,
\eeq
we find that these are given globally by
\begin{align}
2\pi F_0&= N_8^{\mp}=\pm N_8,~~~~-\frac{1}{2\pi}\int_{\mathbb{CP}^1}\hat f_2= N_5 N_8,~~~~\frac{1}{(2\pi)^3}\int_{\mathbb{CP}^2}\hat f_4= \frac{N_5^2N^{\mp}_8}{2},\nn\\[2mm]
-\frac{1}{(2\pi)^5}\int_{\mathbb{CP}^3}\hat f_6&= N_2,~~~~~-\frac{1}{(2\pi)^2}\int_{(r,\mathbb{CP}^1)}H=N_5
\end{align} 
where the $\mp$ superscript indicates that we are on the side of the interior D8 with $r\in \mathbb{R}^{\mp}$ and we have assumed for simplicity that $k=0$ globally in the NS 2-form. 

\begin{figure}
\centering
\includegraphics[scale=0.35]{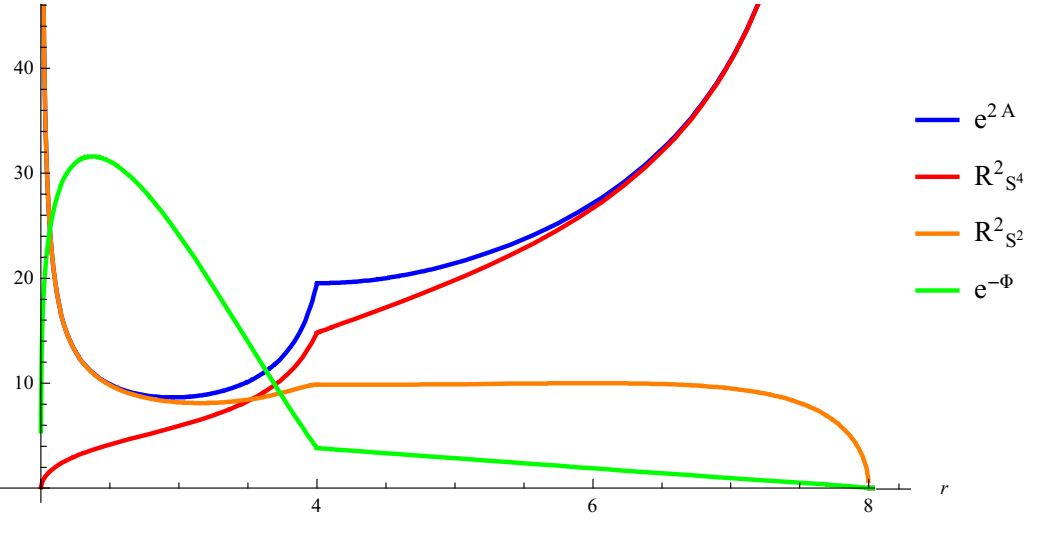}
\caption{Plot of the warp factors in the metric and dilaton for an ${\cal N}=(5,0)$ solution bounded between an O4 plane at $r=2$ and an O6 plane at $r=8$ with a stack of D8 branes at $r= 4$. The remaining parameter is tuned as $b_2=-6$}
\label{fig2}
\end{figure}

With the expressions for the brane charges we can compute the holographic charge via the string frame analogue of the formula presented in \cite{Couzens:2017way}, namely
\beq
c_{hol}=\frac{3}{2^4 \pi^6} \int_{M_7} e^{A-2\Phi}\text{vol}(\text{M}_7),
\eeq
which gives the leading order contribution to the central charge of the putative dual CFT. Given the class of solution is section \ref{ses:classification} we find this expression reduces to 
\beq
c_{hol}= \frac{1}{2} \int \frac{\Delta_1}{u^2} dr.
\eeq
For the case at hand one then finds that
\beq
c_{hol}=N_2 N_5^2 N_8- \frac{3 N_5^5 N_8^2}{20}.
\eeq
The central charge of CFTs with $\mathfrak{osp}(n|2)$ superconformal symmetry takes the form of \eqref{eq:cftcentralcharge}, which in the limit of large level $k$ becomes $c= 3 k$. The holographic central charge is not obviously of this form, however that doesn't mean it is necessarily not the leading contribution to something that is\footnote{Such scenarios are actually quite common, see for instance \cite{Lozano:2019zvg}.}. We leave recovering this result from a CFT computation for future work.\\
~\\
We will now construct a globally bounded solution with interior D8 branes that preserves $\mathcal{N}=(5,0)$ - this time we will be more brief. There are many options for gluing local solutions together for this less supersymmetric case. We will choose to place a D8 brane in one of the bounded behaviour we already found in section \ref{sec:osp52} in the absence of interior D8 branes (see table \ref{table:1}), namely will will insert a D8 in the solution bounded between O6 and O4 places. We remind the reader that we get local solutions containing these singularities by tuning $h$ as
\begin{align}
h_{\text{O4}}&=\frac{1}{2}c_1\big( r_0^2- r_0(r-r_0)+  (r-r_0)^2\big)+\frac{1}{3!}c_2(r-r_0)^3,\nn\\[2mm]
h_{\text{O6}}&=b_1 \tilde{r}_0+b_1 (r- \tilde{r}_0)+\frac{1}{3!}b_2(r-\tilde{r}_0)^3.
\end{align}
where the singularity are located at $(r_0,\tilde{r}_0)$ respectively.  We will assume $r_0,\tilde{r}_0>0$ and place a stack of D8s at a point $r=r_s$ between the two O plane loci. The condition that the NS sector should be continuous in this case amounts to imposing that
\beq
(h_{\text{O4}},h'_{\text{O4}},h''_{\text{O4}})\bigg\lvert_{r=r_s}=(h_{\text{O6}},h'_{\text{O6}},h''_{\text{O6}})\bigg\lvert_{r=r_s},
\eeq
of course we also need the value of $F_0$ to change as we cross the D8. It is indeed possible to solve the continuity condition in this case, which fixes 3 parameters, $(c_1,c_2,b_1)$ say, leaving $(r_0,\tilde{r}_0,b_2)$ as free parameters. A plot of this solution for a choice of  $(r_0,\tilde{r}_0,b_2)$ is given in figure \ref{fig2}.

\section*{Acknowledgements}
We thank Yolanda Lozano, Noppadol Mekareeya and Alessandro Tomasiello for useful discussions. The work of NM is supported by the Ram\'on y Cajal fellowship RYC2021-033794-I, and by grants from the Spanish government MCIU-22-PID2021-123021NB-I00 and principality of Asturias SV-PA-21-AYUD/2021/52177.  AR is partially supported by the INFN grant ``Gauge Theories, Strings and Supergravity'' (GSS).

\appendix
 
\section{Derivation of spinors on $\widehat{\mathbb{CP}}\!\!~^3$ }\label{sec:app1}
In this appendix we derive  all spinors transforming in the \textbf{5} and \textbf{1} of $\mathfrak{so}(5)$ on squashed $\mathbb{CP}^3$. We achieve this by starting with known spinors in the  \textbf{3} of $\mathfrak{so}(3)$ and \textbf{5} of $\mathfrak{so}(5)$ on  the 7-sphere, and then reducing them to $\mathbb{CP}^3$.
\subsection{Killing spinors and vectors on S$^7$= SP(2)/SP(1)}
The 7-sphere admits a parametrisation as an SP(2) bundle over SP(1), ie the SP(2)/SP(1) co-set. For a unit radius 7-sphere this has the  metric
\beq\label{eq:S7met}
ds^2(\text{S}^7)= \frac{1}{4}\bigg[d\alpha^2+ \frac{1}{4}\sin^2 \alpha(L_1^i)^2+ \big(L_2^i- \cos^2 \left(\frac{\alpha}{2}\right) L_1^i\big)^2\bigg]
\eeq 
where we take the following basis of SU(2) Left invariant 1-forms
\beq
L^1_{1,2}+i  L^2_{1,2}= e^{i \psi_{1,2}}(i d\theta_{1,2}+ \sin\theta_{1,2} d\phi_{1,2}),~~~~ L^3_{1,2}= d\psi_{1,2}+ \cos\theta_{1,2} d\phi_{1,2}\label{eq:leftinvariantforms}.
\eeq 
The 7-sphere admits two sets of Killing spinors obeying the relations
\beq\label{eq:7-sphereKSE}
\nabla_{a}\xi_{\pm}= \pm \frac{i}{2}\gamma_{a}\xi_{\pm}.
\eeq
With respect to the vielbein and flat space gamma matrices
\begin{align}\label{eq:frameS7}
e^1&=\frac{1}{2}d\alpha,~e^{2,3,4}=\frac{1}{4}\sin\alpha L^{1,2,3}_1,~e^{5,6,7}= L_2^{1,2,3}- \cos^2 \left(\frac{\alpha}{2}\right) L_1^{1,2,3},\nn\\[2mm]
\gamma_1&=\sigma_1\otimes\mathbb{I}_2\otimes \mathbb{I}_2,~~~
\gamma_{2,3,4}=\sigma_2\otimes\sigma_{1,2,3}\otimes \mathbb{I}_2~~~\gamma_{5,6,7}=\sigma_3\otimes\mathbb{I}_2\otimes \sigma_{1,2,3},
\end{align}
where $\sigma_{1,2,3}$ are the Pauli-matrices, the Killing spinor equation \eqref{eq:7-sphereKSE} is solved by 
\begin{align}\label{eq:squahedS7spinor}
\xi_\pm&={\cal M}_{\pm}\xi^0_{\pm},\\[2mm]
{\cal M}_{\pm}&=e^{\frac{\alpha}{4}(\pm i\gamma_1+Y)}e^{\mp i\frac{\psi_1}{2}\gamma_7P_{\mp}}e^{\mp i\frac{\theta_1}{2}\gamma_6 P_{\mp}}e^{\mp i\frac{\phi_1}{2}\gamma_7 P_{\mp}}e^{\frac{\psi_2}{4}(\pm i\gamma_7+X)}e^{\frac{\theta_2}{2}(\gamma_{13}P_+\pm \gamma_6 P_{\pm})}e^{\frac{\phi_2}{2}(\gamma_{14}P_+\pm i \gamma_7P_{\pm})}\nn,
\end{align}
where $\xi^0_{\pm}$ are unconstrained constant spinors and
\beq
P_{\pm}=\frac{1}{2}(\mathbb{I}_4\pm \gamma_{1234}),~~~X=\gamma_{14}-\gamma_{23}-\gamma_{56},~~~~Y=-\gamma_{25}-\gamma_{36}-\gamma_{47}.
\eeq
It was shown in \cite{Legramandi:2020txf} that $\xi_-$ transform in the $(\textbf{2},\textbf{4})$ of $\mathfrak{sp}(1)\oplus \mathfrak{sp}(2)$ and $\xi_+$ in the $\textbf{3}\oplus \textbf{5}$ of $\mathfrak{so}(3)\oplus\mathfrak{so}(5)$ - it is the latter that will be relevant to us here. Denoting the \textbf{3} and \textbf{5} as $\xi^i_3$ for $i=1,...,3$ and $\xi^{\alpha}_5$ for $\alpha=1,...,5$ and defining the 8 independent supercharges contained in $\xi_+$ as
\beq
\xi^I_{+}= {\cal M_+} \hat\eta^I,~~I=1,...,8,
\eeq 
where the $I^{th}$ entry of $\hat\eta^I$ is 1 and the rest zero,
these are given specifically by
\begin{align}
\hat{\xi}^i_3= \frac{1}{\sqrt{2}}\left(\begin{array}{c}i(-\xi^5_++\xi^8_+)\\
\xi^5_++\xi^8_+\\
i(\xi^6_++\xi^7_+)\\
\end{array}\right)^{i},~~~ \hat{\xi}^{\alpha}_5=\frac{1}{\sqrt{2}} \left(\begin{array}{c}-i(\xi^1_++\xi^4_+)\\
\xi^1_+-\xi^4_+\\
-i(\xi^2_+-\xi^3_+)\\
\xi^2_++\xi^3_+\\
\xi^6_+-\xi^7_+
\end{array}\right)^{\alpha},
\end{align} 
which obey 
\beq
\xi^{i\dag}_3\xi_3^j=\delta^{ij},~~~~\xi^{\alpha \dag}_5\xi_5^{\beta}=\delta^{\alpha\beta},~~~~\xi^{i\dag}_3\xi_5^{\beta}=0,
\eeq
and are Majorana with respect to the intertwiner $B=\sigma_3\otimes \sigma_2\otimes \sigma_2$. The specific Killing vectors that make up the relevant SO(3) and SO(5) in the full space are made up of the following isometries of the base and fibre metrics
\begin{subequations}
\begin{align}
K^{1L}_{1,2}+ i K^{2L}_{1,2}&= e^{i\phi_{1,2}} \left(i \partial_{\theta_{1,2}}+\frac{1}{\sin\theta_{1,2}}\partial_{\psi_{1,2}}-\frac{\cos\theta_{1,2}}{\sin\theta_{1,2}} \partial_{\phi_{1,2}} \right),~~~K^{3L}_{1,2}= -\partial_{\phi_{1,2}},\label{eq:iso1}\\[2mm]
K^{1R}_{1,2}+ i K^{2R}_{1,2}&= e^{i\psi_{1,2}}\left(i \partial_{\theta_{1,2}}+\frac{1}{\sin\theta_{1,2}}\partial_{\phi_{1,2}}-\frac{\cos\theta_{1,2}}{\sin\theta_{1,2}} \partial_{\psi_{1,2}} \right),~~~K^{3R}_{1,2}= \partial_{\psi_{1,2}}\label{eq:iso2},\\[2mm]
\hat{K}^A_{\text{SO(5)/SO(4)}}&= - (\mu_A\partial_{\alpha}+\cot\alpha \partial_{x_i}\mu_A g_3^{ij} \partial_{x_j}), ~~~A=1,...,4 \label{eq:iso3}
\end{align}
\end{subequations}
where $\mu_{A}$ are embedding coordinates for the S$^3\subset$ S$^4$, $g_3^{ij}$ is the inverse metric of this 3-sphere  and $x_i= (\theta_1,\phi_1,\psi_1)_i$, we have specifically
\beq\label{eq:mudef}
\mu_A= \bigg(\sin\left(\frac{\theta_1}{2}\right)\cos\left(\frac{\phi_-}{2}\right),~\sin\left(\frac{\theta_1}{2}\right)\sin\left(\frac{\phi_-}{2}\right),~\cos\left(\frac{\theta_1}{2}\right)\cos\left(\frac{\phi_+}{2}\right),~-\cos\left(\frac{\theta_1}{2}\right)\sin\left(\frac{\phi_+}{2}\right)\bigg)_A,\nn
\eeq
for $\phi_{\pm}=\phi_1\pm\psi_1$. In terms of the isometries on the base and fibre we define the following Killing vectors on the 7-sphere
\beq
K^{iD}=K^{iR}_{1}+K^{iR}_{2},~~~~K^A_{\text{SO(5)/SO(4)}}=\hat{K}^A_{\text{SO(5)/SO(4)}}+ \cot\left(\frac{\alpha}{2}\right)\mu_B(\kappa_{A})^B_i K^{i R}_2.
\eeq
where
\beq
\kappa_1=\left(\begin{array}{ccc}0&0&0\\0&0&-1\\0&-1&0\\-1&0&0\end{array}\right),~~~\kappa_2=\left(\begin{array}{ccc}0&0&1\\0&0&0\\-1&0&0\\0&1&0\end{array}\right),~~~\kappa_3=\left(\begin{array}{ccc}0&1&0\\1&0&0\\0&0&0\\0&0&-1\end{array}\right),~~~\kappa_4=\left(\begin{array}{ccc}1&0&0\\0&-1&0\\0&0&1\\0&0&0\end{array}\right)\nn.
\eeq
The isometry groups in the full space are spanned by
\begin{align}
\text{SO}(3)&:~ K^{iL}_{2}\nn\\[2mm]
\text{SO}(5)&:~(K^{iL}_{1},~K^{iD},~K^A_{\text{SO(5)/SO(4)}})\label{eq:SO(5)}
\end{align}
Another Killing vector on S$^7$ that will be relevant is
\beq
\tilde{K}= y_i K^{iR}_{1},~~~~~y_i=(\cos\psi_2\sin\theta_2,~\sin\psi_2\sin\theta_2,~\cos\theta_2).
\eeq
In terms of this one can define Killing vectors that together with the SO(5) Killing vectors span SO(6), namely
\beq\label{eq:SO(6)}
\text{SO(6)/SO(5)}:~ ([K^A_{\text{SO(5)/SO(4)}},~\tilde {K}],~\tilde {K})
\eeq

\subsection{Reduction to $\mathbb{CP}^3$}
It is possible to rewrite \eqref{eq:S7met} as fibration of $\partial_{\phi_2}$ over $\mathbb{CP}^3$ as
\begin{align}
ds^2(\text{S}^7)&= ds^2(\mathbb{CP}^3)+  \frac{1}{4}\left(d\phi_2+\cos\theta_2 d\psi_2-\cos^2\left(\frac{\alpha}{2}\right) y_i L_1^i\right)^2,\nn\\[2mm]
ds^2(\mathbb{CP}^3)&=\frac{1}{4}\bigg[d\alpha^2+ \frac{1}{4}\sin^2 \alpha(L_1^i)^2+ Dy_i^2\bigg],~~~~Dy_i= dy_i+\cos^2\left(\frac{\alpha}{2}\right)\epsilon_{ijk}y_j  L^k_1.
\end{align}
This can be achieved by rotating the  5,6,7 components of the vielbein in \eqref{eq:frameS7} by
\beq
e^i\to \Lambda^i_{~j}e^j,~~~~\Lambda= \left(\begin{array}{ccc}-\sin\psi_2&\cos\psi_2&0\\
 -\cos\theta_2\cos\psi_2&-\cos\theta_2\sin\psi_2&\sin\theta_2\\ 
\sin\theta_2\cos\psi_2&\sin\psi_2\sin\theta_2&\cos\theta_2\end{array}\right).
\eeq
The corresponding action on the spinors is defined through the matrix
\beq
\Omega= e^{\frac{\theta_2}{2}\gamma_{67}}e^{(\frac{\psi_2}{2}+\frac{\pi}{4})\gamma_{56}}.
\eeq
The \textbf{3} and \textbf{5} in the rotated frame then take the form
\beq
\xi_3^i=\Omega\hat{\xi}^i_3,~~~~ \xi_5^{\alpha}=\Omega\hat{\xi}_5^{\alpha}.
\eeq
Any component of these spinor multiplets that is un-charged under $\partial_{\phi_2}$ is spinor on $\mathbb{CP}^3$, as we have rotated to a frame where translational invariance in $\phi_2$ is manifest, this is equivalent to choosing the parts of $(\xi_3^i,\xi^{\alpha}_5)$ that are independent of $\phi_2$. It is not hard to establish that this is all of $\xi^{\alpha}_5$ and $\xi_3^3$, which being a singlet under SO(5) we now label
\beq
\xi_0= \xi_3^3.
\eeq
The chirality matrix on $\mathbb{CP}^3$ is identified as $\hat\gamma =\gamma_7$, which is clearly an SO(5) singlet, so we can construct an additional SO(5) quintuplet and singlet by acting with this. Additionally
we can define a set of embedding coordinates on S$^4$ via
\beq
i\xi_0^{\dag}\gamma_7\xi^{\alpha}_5= Y_{\alpha},~~~~Y_{\alpha}=(\sin\alpha \mu_A,~\cos\alpha),
\eeq
where $\mu_A$ are embedding coordinates on S$^3$ defined in \eqref{eq:mudef} - as the left hand side of this expression is a quintuplet so too are these embedding coordinates.  In summary we have the following Majorana spinors on $\mathbb{CP}^3$ respecting the $\textbf{5}\oplus \textbf{1}$ branching of SO(6) under its SO(5) subgroup
\beq\label{eq:spinorreps}
\textbf{5}:~ (\xi^{\alpha}_5,~i\gamma_7\xi^{\alpha}_5,~Y_{\alpha}\xi_0,~i Y_{\alpha}\gamma_7\xi_0),~~~~~\textbf{1}:~ (\xi_0,~i\gamma_7 \xi_0),
\eeq
which can be used to construct ${\cal N}=(5,0)$ and ${\cal N}=(1,0)$ AdS$_3$ solutions respectively. One might wonder if one can generate additional spinor in the \textbf{1} or \textbf{5} by acting with the  SO(5) invariant forms one can define on $\mathbb{CP}^3$. These are quoted in the main text in \eqref{eq:J} and \eqref{eq:SO5s},  one can show that on round unit radius $\mathbb{CP}^3$
\begin{align}
\nu^1_2\xi_5^{\alpha}&=-2 Y_{\alpha}\xi_0+i \gamma_7\xi_5^{\alpha},~~\nu^2_2\xi_5^{\alpha}=-2 Y_{\alpha}\xi_0,~~\text{Re}\Omega_3\xi_5^{\alpha}=-4i Y_{\alpha}\xi_0,~~\text{Im}\Omega_3\xi_5^{\alpha}=-4 Y_{\alpha}\gamma_7 \xi_0,\nn\\[2mm]
\nu^1_2\xi_0&=-i\gamma_7 \xi_0,~~\nu^2_2\xi_0=-2i\xi_0,~~\text{Re}\Omega_3\xi_0=-4\gamma_7\xi_0,~~\text{Im}\Omega_3\xi_0=4 i\gamma_7\xi_0,
\end{align}
where $2\nu_2^1=\tilde J_2- J_2$, $2\nu_2^2=\tilde J_2+J_2$ and the forms should be understood as acting on the spinors  through the Clifford map  $X_n\to \frac{1}{n!}(X_n)_{a_1...a_n}\gamma^{a_1...a_n}$, so \eqref{eq:spinorreps} are in fact exhaustive. The SO(6) Killing vectors on $\mathbb{CP}^3$ are given by \eqref{eq:SO(5)} and \eqref{eq:SO(6)} but with the $\partial_{\phi_2}$ dependence omitted, the SO(5) vectors are still Killing when one allows the base S$^4$ to have a different radi to the S$^2$ in the $\mathbb{CP}^3$ metric (ie for squashed $\mathbb{CP}^3$) this however breaks the SO(6)/SO(5) isometry. Finally note that $\xi_0$ are actually charged under SO(6)/SO(5), and more specifically when these isometries are not broken (ie $\mathbb{CP}^3$ is not squashed) then we have the following  independent SO(6) sextuplets
\beq\label{eq:the6}
\textbf{6}:~\xi^{\cal I}_6=\left(\begin{array}{c}\xi_5^{\alpha}\\\xi_0\end{array}\right)^{\cal I},~~~~ \hat\xi^{\cal I}_6=\left(\begin{array}{c}i\gamma_7\xi_5^{\alpha}\\i\gamma_7\xi_0\end{array}\right)^{\cal I},
\eeq
which can be used to construct ${\cal N}=(6,0)$ AdS$_3$ solutions.

\section{The SO(3)$_L\times$SO(3)$_D$ invariant ${\cal N}=5$ bi-linears}\label{sec:bilinears}
In the main text we will need to construct ${\cal N}=5$ bi-linears on the space \eqref{eq:defCP3squahsed}, the non trivial part of this computation comes from the bi-linears on squashed $\mathbb{CP}^3$ - in this appendix we shall compute them.\\
~~\\
As explained in the main text it is sufficient to solve the supersymmetry constraints for an ${\cal N}=1$ sub-sector of the quintuplet of SO(5) spinors defined on the internal space. A convenient component to work with is the 5th as this is a singlet under an SO(4) subgroup of SO(5). Specifically with respect to \eqref{eq:defCP3squahsed} and the discussion below it, $\chi^5_{1,2}$ are singlets with respect to SO(4)=SO(3)$_L\otimes$SO(3)$_D$. As such the bi-linears that follow from $\chi^5_{1,2}$ must decompose in a basis of the SO(3)$_L\otimes$SO(3)$_D$ invariant forms on the S$^2\times$S$^3$ fibration \eqref{eq:SO3Dinvariants} 
and what one can form from these through taking wedge products. The $d=7$ spinors $\chi^5_{1,2}$ depend  on $\hat{\mathbb{CP}}^3$ through 
\beq
\eta_{\pm}= \xi^{5}_5\pm i Y_5 \hat\gamma \xi_0,~~~~ Y_5 \xi_0,~~~~Y_5 i\hat\gamma \xi_0
\eeq
where these are all defined in the previous appendix - it is the bi-linears we can construct out of these that will be relevant to us. One can show that 
\begin{align}
\eta_{\pm}\otimes \eta_{\pm}^{\dag}&= \phi^1_+\pm i \phi^1_-,~~~~~\eta_{\pm}\otimes \eta_{\mp}^{\dag}= \pm\phi^2_+ +i \phi^2_-,\nn\\[2mm]
\eta_{+}\otimes \xi_{0}^{\dag}&=\phi^3_++ i \phi^3_-,~~~~~\eta_{-}\otimes (i\hat\gamma\xi_{0})^{\dag}=\phi^3_+- i \phi^3_-,\nn\\[2mm]
\eta_{+}\otimes (i\hat\gamma\xi_{0})^{\dag}&=\phi^4_++i \phi^4_-,~~~~\eta_{-}\otimes \xi_{0}^{\dag}=-\phi^4_++i \phi^4_-,\nn\\[2mm]
\xi_{0}\otimes \eta_+^{\dag}&=\phi^5_++ i \phi^5_-,~~~~~~i\hat\gamma\xi_{0}\otimes \eta_-^{\dag}=\phi^5_+- i \phi^5_-,\nn\\[2mm]
i\hat\gamma \xi_0\otimes \eta_+^{\dag}&= \phi^6_++ i \phi^6_-,~~~~~~\xi_0\otimes \eta_-^{\dag}=-\phi^6_++ i \phi^6_-,\nn\\[2mm]
\xi_0\otimes \xi_0^{\dag}&=\phi^7_++ i \phi_-^7,~~~~~i\hat\gamma\xi_0\otimes (i\hat\gamma\xi_0)^{\dag}=\phi^7_+- i \phi_-^7,\nn\\[2mm]
i \hat\gamma \xi_0\otimes \xi_0^{\dag}&= \phi^8_++ i \phi^8_-,~~~~~\xi_0\otimes (i \hat\gamma \xi_0)^{\dag}=- \phi^8_++ i \phi^8_-,
\end{align}
where $\phi^{1...8}_{\pm}$  are real bi-linears of even/odd form degree, they take the form
\begin{align}
\phi^1_+&=\frac{1}{8}\sin^2\alpha\bigg(1- \frac{1}{32}e^{2(C+D)}\sin^2\alpha\omega_2^2\wedge \omega_2^2+\frac{1}{16} e^{2C}\sin\alpha d\alpha\wedge \omega_1\wedge \big(e^{2D} \omega_2^1-e^{2C}\sin^2\alpha \omega^4_2\big)\bigg),\nn\\[2mm]
\phi^1_-&=\frac{1}{64}e^{2C+D}\sin^4\alpha \omega_1\wedge \omega^2_2,~~~~\phi^2_-=-\frac{1}{64}e^{2C+D}\sin^3\alpha\bigg(\sin\alpha\omega_1\wedge \omega^3_2+  d\alpha \wedge \omega^2_2\bigg)\nn\\[2mm]
\phi^2_+&=\frac{1}{32}\sin^2\alpha\bigg(e^{2C}\sin^2\alpha\omega^4_2-e^{2D}\omega^1_2+ e^{2C}\sin\alpha d\alpha\wedge \omega_1\wedge\big(1-\frac{1}{32}e^{2(C+D)}\sin^2\alpha \omega_2^2\wedge \omega_2^2\big) \bigg),\nn\\[2mm]
\phi^3_+&=\frac{1}{32}\sin^2\alpha \bigg(-e^{C+D}\omega^2_2+ \frac{1}{4}e^{3C+D}\sin\alpha d\alpha \wedge \omega_1\wedge \omega_2^3\bigg),\nn\\[2mm]
\phi^3_-&=\frac{1}{16}\sin\alpha \bigg(e^C\sin\alpha \omega_1\wedge\big(1-\frac{1}{32}e^{2(C+D)}\sin^2\alpha \omega_2^2\wedge \omega_2^2\big)+\frac{1}{4}d\alpha\wedge \big(e^{C+2D}\omega^1_2- e^{3C}\sin^2\alpha \omega_2^4\big)\bigg),\nn\\[2mm]
\phi^4_+&=-\frac{1}{32}e^{C+D}\sin^2\alpha\bigg(\omega^3_2+ \frac{1}{4}e^{2C}\sin\alpha d\alpha\wedge \omega_1 \wedge \omega^2_2\bigg),~~~~\phi^5_+=\frac{1}{32}e^{C+D}\sin^2\alpha\bigg(\omega^2_2+\frac{1}{4}e^{2C}\sin\alpha d\alpha\wedge \omega_1 \wedge \omega^3_2\bigg),\nn\\[2mm]
\phi^4_-&= \frac{1}{16}e^{C}\sin\alpha\bigg(\frac{1}{4}\sin\alpha \omega_1\wedge \big(e^{2D}\omega_2^1-e^{2C}\sin^2\alpha \omega_2^4\big)-d\alpha\wedge\big(1-\frac{1}{32}e^{2(C+D)}\sin^2\alpha\wedge \omega^2_2\wedge \omega^2_2\big)\bigg),\nn\\[2mm]
\phi^5_-&=\frac{1}{16}e^C \sin\alpha\bigg(\frac{1}{4}d\alpha\wedge \big(e^{2D}\omega^1_2- e^{2C}\sin^2\alpha\omega_2^4\big)-\sin\alpha \omega_1\wedge \big(1-\frac{1}{32}e^{2(C+D)}\sin^2\alpha \omega_2^2\wedge \omega_2^2\big)\bigg),\nn\\[2mm]
\phi^6_+&=\frac{1}{32}e^{C+D}\sin^2\alpha\bigg(\omega^3_2-\frac{1}{4}e^{2C}\sin\alpha d\alpha\wedge \omega_1 \wedge \omega_2^2\bigg),\nn\\[2mm]
\phi^6_-&=\frac{1}{16}e^{C}\sin\alpha\bigg(d\alpha\wedge \big(1-\frac{1}{32} e^{2(C+D)}\sin^2\alpha \omega^2_2\wedge \omega^2_2\big)+\frac{1}{4}\sin\alpha \big(e^{2D}\omega^1_2-e^{2C}\sin^2\alpha\omega^4_2\big)\bigg),\nn\\[2mm]
\phi^7_+&=\frac{1}{8}\bigg(1+\frac{1}{16}e^{2C}\sin\alpha d\alpha\wedge \omega_1\wedge\big(-e^{2D}\omega^1_2+e^{2C}\sin^2\alpha \omega^4_2\big)-\frac{1}{32}e^{2(C+D)}\sin^2\alpha \omega^2_2\wedge \omega^2_2\bigg),\nn\\[2mm]
\phi^7_-&=\frac{1}{64}e^{2C+D}\sin\alpha\bigg(d\alpha\wedge \omega^3_2+\sin\alpha \omega^1\wedge \omega^2_2\bigg),~~~~\phi^8_-= \frac{1}{64}e^{2C+D}\sin\alpha\bigg(-d\alpha\wedge \omega^2_2+\sin\alpha \omega_1\wedge \omega^3_2\bigg)\nn\\[2mm]
\phi^8_+&=-\frac{1}{32}\bigg(e^{2C}\sin\alpha d\alpha\wedge \omega_1\wedge \big(1-\frac{1}{32}e^{2(C+D)}\sin^2\alpha \omega_2^2\wedge \omega_2^2\big)+ e^{2D}\omega^1_2-e^{2C}\sin^2\alpha \omega^4_2\bigg)\label{eq:SO3Dpolyforms}
\end{align}

\section{Ruling out $\mathfrak{osp}(7|2)$ AdS$_3$ vacua}\label{app:sop72nogo}
In this appendix we shall prove that all ${\cal N}=7$ AdS$_3$ solutions preserving the algebra $\mathfrak{osp}(7|2)$ are locally AdS$_4\times$S$^7$.\\
~~\\
$\mathfrak{osp}(7|2)$ necessitates an SO(7) R-symmetry with spinor transforming in the \textbf{7}, there is only one way to achieve this. On needs a  round 7-sphere in the metric with fluxes that break its SO(8) isometry to SO(7) in terms of the weak G$_2$ structure 3-forms one can define. Such an ansatz in type II can be ruled out a the level of the equations of motion \cite{Legramandi:2020txf}, our focus here then will be on $d=11$ supergravity.\\
~~\\
All AdS$_3$ solutions of 11 dimensions supergravity admit a decomposition of their  bosonic fields as
\beq\label{eq:ansatzMtheory}
ds^2= e^{2A} ds^2(\text{AdS}_3)+ ds^2(\text{M}_8),~~~~G= e^{3A}\text{vol}(\text{AdS}_3)\wedge F_1+ F_4
\eeq 
where $F_1,F_4,A$ have support on $\text{M}_8$ only. We take AdS$_3$ to have inverse radius $m$.  When a solutions is supersymmetric M$_8$ supports (at least one) Majorana spinor $\chi$ that one can use to define the following bi-linears
\begin{align}\label{eq:mtheoryforms}
2e^A&= |\chi|^2,~~~2e^Af= \chi^{\dag}\hat\gamma^{(8)}\chi,~~~~2 e^{A}K =\chi^{\dag}_+\gamma^{(8)}_a\chi_- e^a ,\nn\\[2mm]
2e^A\Psi_3& = \frac{1}{3!}\chi^\dag\gamma^{(8)}_{abc}\hat\gamma^{(8)}\chi e^{abc},~~~2e^A\Psi_4=\frac{1}{4!}\chi^\dag\gamma^{(8)}_{abcd}\chi e^{abcd}
\end{align}
where $\gamma^{(8)}_a$ are eight-dimensional flat space gamma matrices, $\hat\gamma^{(8)}=\gamma^{(8)}_{12345678}$ is the chirality matrix and $e^a$ is a vielbein on M$_8$. Sufficient conditions for ${\cal N}=1$ supersymmetry to hold can be caste as the following  differential conditions the bi-linears should obey \cite{Legramandi:2020txf}
\begin{subequations}
\begin{align}
& d(e^{2A} K)=0,\label{eq:MtheoryBPS1}\\[2mm]
&d(e^{3A} f)- e^{3A} F_1-2 m  e^{2A} K=0,\label{eq:MtheoryBPS2}\\[2mm]
& d(e^{3A} \Psi_3)- e^{3A}(-\star_8 F_4+ f F_4)+2 m e^{2A} \Psi_4=0\label{eq:MtheoryBPS3},\\[2mm]
&d (e^{2A} \Psi_4)- e^{2A} K \wedge F_4=0\label{eq:MtheoryBPS4},\\[2mm]
&6 \star_8 dA-2 f \star_8 F_1+ \Psi_3\wedge F_4=0,\label{eq:MtheoryBPS5}\\[2mm]
&6 e^{-A} m \star_8 K - 6 f \star_8 dA+2 \star_8 F_1+\Psi_3 \wedge \star_8 F_4=0\label{eq:MtheoryBPS6}
\end{align}
\end{subequations}
where $\star_8$ is the hodge dual on the M$_8$. These conditions do not imply all of the equations of motion of 11 dimensional supergravity however. For that to follow one must additionally solve the Bianchi identity and equation of motion of the 4-form flux $G$. Away from the loci of sources, this amounts to imposing that 
\beq\label{eq:BianchiMtheory}
d(F_4)=0,~~~~ d(\star_8 F_1)-\frac{1}{2} F_4\wedge F_4=0.
\eeq
The only way to realise the SO(7) R-symmetry that $\mathfrak{osp}(7|2)$ necessitates on a 8d space is to take it to be a foliation of the SO(7)/G$_2$ co-set over an interval.  As explained at greater length in section 6.2 of \cite{Legramandi:2020txf}, the metric on this co-set is the round one, but the flux can depend also depend on a SO(7) invariant 3-form $\phi^0_3$ such that \eqref{eq:ansatzMtheory} should be refined as
\begin{align}\label{eq:SO(7)ansatz}
ds^2(\text{M}_8)&= e^{2B}ds^2(\text{S}^7)+ e^{2k}dr^2, ~~~~e^{3A}F_1= f_1 dr,~~~F_4=  4f_2 \star_{7}\phi^0_3+ f_3 dr\wedge\phi^0_3.  
\end{align}
where $(e^{2A},e^{2B},e^{2k},f_i)$ are functions of the interval only.  The  SO(7) invariants obey the following relations
\beq
d\phi^0_3=4 \star_7\phi^0_3,~~~\phi^0_3\wedge \star_7\phi^0_3=7 \text{vol}(\text{S}^7),
\eeq
ie they define the structure of a manifold of weak G$_2$ holonomy. More specifically, decomposing 
\beq
ds^2(\text{S}^7)=d\alpha^2+\sin^2\alpha ds^2(\text{S}^6)
\eeq
One has
\begin{align}
\phi^0_3&= \sin^2\alpha d\alpha\wedge J_{\text{G}_2}+\sin^3\!\alpha\,\text{Re}(e^{-i \alpha}\Omega_{\text{G}_2}),~~~\star_7\phi^0_3=-\frac{1}{2}\sin^4\!\alpha\,J_{\text{G}_2}\wedge J_{\text{G}_2}-\sin^3\!\alpha \,d\alpha\wedge\text{Im}(e^{-i \alpha}\Omega_{\text{G}_2}),\nn\\[2mm]
J_{\text{G}_2}&=\frac{1}{2}{\cal C}_{ijk}Y_{\text{S}^6}^{i}dY_{\text{S}^6}^{j}\wedge dY_{\text{S}^6}^{k},~~~\Omega_{\text{G}_2}=\frac{1}{3!}(1-i \iota_{d\alpha}\star_6){\cal C}_{ijk}dY_{\text{S}^6}^{i}\wedge dY_{\text{S}^6}^{j}\wedge dY_{\text{S}^6}^{k},
\end{align}
where $Y_{\text{S}^6}^{i}$ are unit norm embedding coordinates for $\text{S}^6$ and ${\cal C}_{ijk}$ are the structure constants defining the product between the octonions, ie $o^io^j=-\delta^{ij}+\mathcal{C}^{ijk}o_k$. 
The Killing spinors on unit radius S$^7$ obeying the equation
\beq
\nabla^{(7)}_a\xi=\frac{i}{2}\gamma^{(7)}_a\xi,
\eeq
branch as $\textbf{1}+\textbf{7}$ under the SO(7) subgroup of SO(8), we denote the portions of $\xi$ that transform in these reps as respectively $\xi^0$ and $\xi^I_{\textbf{7}}$, they can be extracted from the relations
\beq
\textbf{1}:~(\phi^0_3+\frac{i}{7})\xi=0,~~~\textbf{7}:~(\phi^0_3- i)\xi=0,
\eeq
where both the \textbf{1} and \textbf{7} are Majorana. Acting with the SO(7) invariants on $\xi^I_{\textbf{7}}$ does not generate any additional spinors in the \textbf{7}, and we can without loss of generality take
\beq
|\xi^0|^2=|\xi_{\textbf{7}}^I|^2=1,~~~ \xi^{0\dag}\xi_{\textbf{7}}^I=0.
\eeq
Thus we only have 1 spinor in the \textbf{7} and the most general Majorana spinors we can write on M$_8$ are $\chi= \sqrt{2}e^{\frac{A}{2}}(\chi_++\chi_-)$ where\footnote{We define the 8d gamma matrices as $\gamma^{(8)}_a=\sigma_1\otimes \gamma^{(7)}_a$ for $a=1,..,7$ and  $\gamma^{(8)}_8=\sigma_2\otimes \mathbb{I}$ where the intertwiner defining Majorana conjugation is $B^{(8)}=\sigma_3\otimes B^{(7)}$. This is the reason for the form that the interval components of the spinors take. }
\beq
\chi_+=\left(\begin{matrix}a_+\\ 0 \end{matrix}\right)\otimes \xi_{\textbf{7}}^I,~~~\chi_+=\left(\begin{matrix}0\\ i a_- \end{matrix}\right)\otimes \xi_{\textbf{7}}^I
\eeq
where $a_{\pm}$ are real functions subject to $|a_+|^2+|a_-|^2=1$ - which are clearly rather constrained.  The bi-linears of each component of $\xi_{\textbf{7}}^I$ give rise to another 7  weak G$_2$ holonomy 3-forms as
\beq
\xi^{(I)}_{\textbf{7}}\otimes \xi^{(I)\dag}_{\textbf{7}}=\frac{1}{8}\big(1+i \phi_3^{(I)}+\star_7 \phi_3^{(I)}+\text{vol}(\text{S}^7)\big).
\eeq
As $\phi_3^{(I)}$ are charged under SO(7) they are clearly all independent of  $\phi^{0}_3$, so there is no way to generate the invariant forms in the flux in \eqref{eq:SO(7)ansatz} from \eqref{eq:MtheoryBPS3}-\eqref{eq:MtheoryBPS6}, thus we must have
\beq
f_2=f_3=0,~~~~\Rightarrow~~~ F_4=0.
\eeq
This makes the flux purely electric and it is proved in \cite{Martelli:2003ki}, that for all such solutions AdS$_3$ experiences an enhancement to AdS$_4$. As there is no longer anything breaking the isometries of the 7-sphere locally, clearly then this ansatz just leads to local AdS$_4\times$S$^7$ . The only global possibility beyond the standard ${\cal N}=8$ M2 brane near horizon is  an orbifolding of the 7-sphere that breaks supersymmetry to  ${\cal N}=7$ - in any case this is certainly in no way an AdS$_3$ vacuum.

\end{document}